\documentclass[11pt]{article}

\usepackage[T1]{fontenc}
\usepackage{lmodern}
\usepackage[margin=1in]{geometry}
\usepackage{amsmath,amsfonts,amsthm}
\usepackage{microtype}
\usepackage[colorlinks=true,linkcolor=blue,citecolor=blue,urlcolor=blue]{hyperref}
\usepackage[capitalise,noabbrev,nameinlink]{cleveref}

\newtheorem{theorem}{Theorem}[section]
\newtheorem{lemma}[theorem]{Lemma}
\newtheorem{corollary}[theorem]{Corollary}
\theoremstyle{definition}

\numberwithin{equation}{section}

\DeclareMathOperator{\poly}{poly}
\newcommand{\MC}{\operatorname{MaxCut}}
\newcommand{\Cut}{\operatorname{Cut}}
\newcommand{\E}{\mathbb{E}}
\newcommand{\Prb}{\mathbb{P}}
\newcommand{\R}{\mathbb{R}}
\DeclareMathOperator{\conv}{conv}
\DeclareMathOperator{\OPT}{OPT}
\DeclareMathOperator{\hOPT}{\widehat{\mathrm{OPT}}}
\newcommand{\tO}{\widetilde{O}}

\usepackage{xcolor}

\title{Optimal Sample Complexity for Max-Cut}

\author{Itamar Bender
    \thanks{Email: \texttt{itamar.bender@weizmann.ac.il}
    }
    \qquad
    Robert Krauthgamer
        \thanks{The Harry Weinrebe Professorial Chair of Computer Science.
            Work partially supported by the Israel Science Foundation grant \#1336/23.
            Email: \texttt{robert.krauthgamer@weizmann.ac.il}
        }
    \\  Weizmann Institute of Science
}

\begin{document}
\maketitle

\begin{abstract}
We determine the optimal sample size for constructing value coresets for Max-Cut via
uniform vertex sampling.
That is, sampling $O(\varepsilon^{-2})$ vertices uniformly at random
in $n$-vertex graphs with edge weights in $[0,1]$
yields a reweighted subgraph whose Max-Cut value
is an additive $\varepsilon n^2$-approximation to that of the original graph,
with constant success probability.
This improves the previously known $O(\varepsilon^{-4})$ upper bound
and matches the known lower bound.
The same guarantee holds also for importance sampling,
where vertices are sampled proportionally to their weighted degrees,
and we further prove a matching lower bound for this sampling method.

Our analysis extends directly to metric Max-Cut,
showing that importance sampling of $O(\varepsilon^{-2})$ vertices
yields $(1\pm\varepsilon)$-approximation to the Max-Cut value,
and to Max-Cut in graphs with edge weights in $[0,1]$ and average weighted degree $\bar d$,
achieving the same $(1\pm\varepsilon)$-approximation
by either uniform or importance sampling of $O(\varepsilon^{-2}n/\bar d)$ vertices.
We additionally show how to use this coreset to recover 
a cut of expected value at least $\mathrm{OPT}-\varepsilon n^2$,
by making $O(n\varepsilon^{-2})$ additional edge-weight queries.
Finally, our result extends to Max-$r$-Cut,
giving an additive $\varepsilon n^2$-approximation using
a sample of $O(\varepsilon^{-2}\log(2r))$ vertices.
\end{abstract}

\section{Introduction}
The Maximum Cut problem (Max-Cut) is a fundamental optimization problem that asks for a bipartition
of a graph's vertices that maximizes the total weight of edges crossing between the two
parts.
 Formally, given an undirected graph $G$ with vertex set $V$ and nonnegative edge
weights $A_{uv}$, let
\[
  \mathrm{OPT}=\MC(G)
    :=\max_{S\subseteq V}\sum_{u\in S,\ v\notin S}A_{uv}.
\]
By convention, $A_{uv}=0$ when $(u,v)$ is not an edge,
and all graphs have no self-loops, which do not affect the value of a cut.

The problem is NP-hard and
APX-hard even on unweighted graphs~\cite{karp72,py91}, so a polynomial-time approximation scheme would imply $\mathrm{P}=\mathrm{NP}$. For general graphs, a common goal
is therefore to design an $\alpha$-approximation algorithm for $0<\alpha\le1$, i.e., one
that returns a cut of value at least $\alpha\,\mathrm{OPT}$.
The algorithm of Goemans and Williamson~\cite{gw95} achieves approximation ratio $\approx 0.87856$ in
expectation.
On the hardness side, H{\aa}stad~\cite{hastad01} proved that achieving a ratio
of $16/17+\gamma$ is NP-hard for every fixed $0<\gamma<1/17$.
The ratio achieved by Goemans and Williamson is optimal assuming the Unique Games Conjecture~\cite{kkmo07}.

Dense and metric instances, however, admit $(1-\varepsilon)$-approximations for every
$0<\varepsilon<1$ through additive approximation.
 For graphs with weights
$A_{uv}\in[0,1]$, an additive approximation means finding a cut of value at least
$\mathrm{OPT}-\varepsilon n^2$, where throughout $n:=|V|$.
We call a graph dense if its total edge weight $W:=\frac12\sum_{u,v\in V}A_{uv}$
satisfies $W\ge\rho n^2$ for a fixed $\rho>0$.
It is well known that
$\mathrm{OPT}\ge W/2\ge\rho n^2/2$.
Hence, additive approximation of $\varepsilon\rho n^2/2$
gives a cut of value at least $(1-\varepsilon)\mathrm{OPT}$,
i.e., achieves $(1-\varepsilon)$-approximation.
The first approximation schemes for dense unweighted Max-Cut that run in polynomial time were given 
by Fernandez de la Vega~\cite{vega96} and independently
by Arora, Karger, and Karpinski~\cite{akk99}.
The same additive viewpoint applies also to
metric Max-Cut, where the weights satisfy the triangle inequality. Fernandez de la Vega and
Kenyon~\cite{vegakenyon01} reduce these instances to dense weighted graphs, such that
accurate additive approximation on the resulting dense graph gives a
$(1-\varepsilon)$-approximation for the original metric instance.

We focus on fast additive approximation of the Max-Cut value without reporting a cut. 
That is, for bounded weights $A_{uv}\in[0,1]$, we seek a randomized estimate $\widehat{\mathrm{OPT}}$ that satisfies
\[
  |\widehat{\mathrm{OPT}}-\mathrm{OPT}|\le\varepsilon n^2,
\]
where throughout, randomized guarantees hold with probability at least $2/3$,
unless mentioned otherwise.

\subsection{Sublinear algorithms for Max-Cut}

To obtain the desired additive approximation, it suffices to sample vertices uniformly and
examine their induced subgraph. The number of sampled vertices depends only on
$\varepsilon$~\cite{afkk03,rv07}.
 The estimator simply computes Max-Cut on this graph after
rescaling all edge weights by a suitable factor. 
Such a small reweighted graph is a
\emph{value coreset} because its Max-Cut value approximates that of the original graph.
With access to each edge weight in constant time and a fixed target success probability, constructing the value coreset takes time
$\poly(1/\varepsilon)$, independent of $n$.
 The same algorithm is useful also for processing a graph stream. Assuming the vertex set is known in
advance, one can choose the sample before the stream begins and maintain its induced subgraph as updates arrive.

While this sampling algorithm is straightforward, analyzing it is nontrivial because the sampled graph's optimal cut is determined after the sample is realized
and need not correspond to any original cut fixed in advance.
Several different approaches have led to an $\widetilde O(\varepsilon^{-4})$
sample bound.
\footnote{The notation $\tO(f)$ hides logarithmic factors in $f$.}
Alon, Fernandez de la Vega, Kannan, and Karpinski~\cite{afkk03} use matrix
decompositions and linear programming. 
Rudelson and Vershynin~\cite{rv07} use tools from geometric functional analysis, 
whereas Mathieu and Schudy~\cite{ms08} use a greedy construction for unweighted graphs. 
These last two analyses remove the logarithmic factors, 
but retain the $\varepsilon^{-4}$ dependence.

This may be viewed as an indication that a smaller sample bound is not likely.  
For metric Max-Cut, importance sampling proportionally to the weighted degrees 
gives a $(1\pm\varepsilon)$-approximate value coreset 
using $O(\varepsilon^{-4})$ vertices~\cite{cjk23}. 

Currently there is a gap between this $O(\varepsilon^{-4})$ upper bound 
and the $\Omega(\varepsilon^{-2})$ lower bound for uniform sampling proved by
Alon, Fernandez de la Vega, Kannan, and Karpinski~\cite[technical report version]{afkk03}.  

These results raise a few questions about the size of value coresets:
For graphs with weights in $[0,1]$, what is the optimal sample size for
uniform sampling to achieve additive approximation $\varepsilon n^2$?
For metric Max-Cut, can the $O(\varepsilon^{-4})$ sample bound for importance sampling
be improved? In either setting, can a different sampling distribution
reduce the sample size?

\subsection{Our results}

Our main contribution is to close the sampling gap,
by showing that sampling $O(\varepsilon^{-2})$ uniformly random vertices
yields a value coreset for Max-Cut in a graph with weights in $[0,1]$.
This improves the previous $O(\varepsilon^{-4})$ upper
bound~\cite{rv07} and matches the known lower bound~\cite{afkk03}.

\begin{theorem}[Uniform sampling for Max-Cut, follows from \Cref{thm:main}]
\label{thm:mainintro}

Consider an $n$-vertex graph with edge weights $A_{uv}\in[0,1]$.
Sample $k$ vertices $X_1,\ldots,X_k$ uniformly with replacement,
\footnote{In some prior work the sampling is without replacement~\cite{afkk03,ms08}, 
while other work allows replacement~\cite{ggr98}.
For sufficiently large $n$, the difference is negligible;
for example, the birthday paradox calculation bounds 
the probability of a repeated vertex among $k$ uniform draws by $\binom{k}{2}/n$.
}
and let the sampled subgraph on vertex set $[k]$ have edge weights
$\widehat A_{ij}=(n/k)^2 A_{X_iX_j}$.
Then
\[
  \E\bigl| \hOPT -\OPT \bigr|
    =O\!\left(\frac{n^2}{\sqrt{k}}\right) ,
  \qquad
  \text{ where }
  \OPT:=\MC(A),\  
  \hOPT:=\MC(\widehat A) .  
\]
\end{theorem}

Our analysis applies more generally to independent draws from any 
sampling distribution $p\in\mathbb R^V$ satisfying $p_v\ge d(v)/n^2$, 
where $d(v)$ is the weighted degree of $v$;
this includes also importance sampling. 
The same bound on expected error holds after
reweighting the sampled edges as
\[
  \widehat A_{ij}=\frac{A_{X_iX_j}}{k^2p_{X_i}p_{X_j}}.
\]

\begin{theorem}[General sampling distributions, follows from Corollary~\ref{cor:tail}]
\label{thm:generalintro}

Consider an $n$-vertex undirected graph with edge weights $A_{uv}\in[0,1]$, 
and let $p\in\mathbb R^V$ be a sampling distribution satisfying 
$p_v\ge d(v)/n^2$ for every vertex $v$, where $d(v):=\sum_u A_{vu}$.
Then for every $\varepsilon,\delta\in (0,1)$ and a suitable absolute constant $C>0$, 
a reweighted sample of $k\ge C\varepsilon^{-2}\delta^{-2}$ vertices satisfies
\[
  \Prb\!\left(
    |\widehat{\mathrm{OPT}}-\mathrm{OPT}|\le\varepsilon n^2
  \right)\ge1-\delta.
\]
\end{theorem}

We also prove a matching lower bound of
$\Omega(\varepsilon^{-2})$ vertices for importance sampling, even for unweighted graphs
(Section~\ref{sec:importance-lower-bound}).
Together with the known lower bound for uniform sampling~\cite{afkk03}, 
this establishes that the optimal sample complexity is 
$\Theta(\varepsilon^{-2})$ 
for both sampling methods in graphs with bounded weights.

\paragraph{Extensions.} 
We actually prove a more general form of Theorem~\ref{thm:generalintro}
(see Theorem~\ref{thm:main}), 
and it has two additional consequences.

The first one is that in graphs with edge weights in $[0,1]$ and average weighted degree $\bar d$,
either uniform or importance sampling of $O(\varepsilon^{-2} n/\bar d)$ vertices
gives a $(1\pm\varepsilon)$-approximate value coreset 
(see Corollary~\ref{cor:average-degree-relative-error}).
This improves over the previous $\tO(\varepsilon^{-4} n/\bar d)$ bound
of Bhaskara, Daruki, and Venkatasubramanian~\cite{bdv18},
by matching their $n/\bar d$ dependence, 
improving their $\varepsilon^{-4}$ to $\varepsilon^{-2}$,
and removing their $\log n$ factor 
(although they get higher success probability $1-1/n^2$). 

The second consequence is that in metric Max-Cut instances,
importance sampling of $O(\varepsilon^{-2})$ vertices
gives a $(1\pm\varepsilon)$-approximate value coreset
(see Corollary~\ref{cor:metric-relative-error}).
Our direct proof avoids the roundabout analysis through a reduction
to dense graphs used in earlier proofs~\cite{vegakenyon01,cjk23}.

The value coreset also supports \emph{cut recovery},
i.e., finding a nearly-optimal cut of the original graph $G$
(see Section~\ref{sec:solution-recovery}).
For weights in $[0,1]$ and either sampling method,
an optimal cut of the reweighted subgraph
and the edge weights between the sampled vertices
and the other vertices of $G$ define a linear program (LP).
An optimal solution of it associates each vertex of $G$
with a probability to lie on one side of the cut,
and assigning the vertices independently according to these probabilities
gives a cut of expected value at least 
$\mathrm{OPT}-\varepsilon n^2$.

The LP has $n+k$ variables and $2k$ constraints,
where $k=O(\varepsilon^{-2})$ is the sample size,
and constructing it requires $O(n\varepsilon^{-2})$ additional queries to the matrix of edge weights. 

Earlier work on cut recovery for unweighted graphs
by Goldreich, Goldwasser, and Ron~\cite{ggr98}
constructs a cut achieving additive $\varepsilon n^2$-approximation using
$O(\varepsilon^{-7}\log^2(2/\varepsilon)+n\varepsilon^{-2}\log(2/\varepsilon))$
matrix queries.
Yaroslavtsev~\cite{yar14} later obtained a cut achieving an
additive $\varepsilon n^2$-approximation in total running time
$O(n\varepsilon^{-2})+2^{O(\varepsilon^{-2})}$.
Neither algorithm uses linear programming,
although other algorithms for finding $(1-\varepsilon)$-approximate
cuts do~\cite{akk99,flp16,yz14}.

Finally, our result extends to Max-$r$-Cut, 
which asks to partition the vertices into at most $r$ parts
so as to maximize the total weight of edges between different parts
(see Theorem~\ref{thm:rway-main}).
For graphs with edge weights in $[0,1]$ and integer $r\ge2$, either sampling method
achieves additive $\varepsilon n^2$-approximation using
\(O(\varepsilon^{-2}\log(2r))\) sampled vertices.
In comparison with the $\widetilde O(r^8\varepsilon^{-4})$ bound
implied by Karpinski and Mark\'o~\cite{km16},
we improve the dependence on both $r$ and the accuracy parameter $\varepsilon$.

\subsection{Our techniques}
\label{sec:intro-techniques}

Each technical section includes an overview of its proof,
and here we only briefly summarize the main ideas.
Our proof for Max-Cut in Section~\ref{sec:max-cut-sample-complexity}
expresses the (normalized) weight of cuts in the graph
as a quadratic form in a vector $p\in\R^V$ that represents the cut.
We compare the maxima of this quadratic form over two feasible regions:
one determined by the sampling distribution (e.g., uniform over $V$),
and the other by the empirical distribution of the sample.
We relax the quadratic objective to a family of linear objectives by letting
the two vector factors vary independently and optimizing over one with the other fixed.
A minimax argument shows that if the linear objectives have nearly the same
maximum over the two feasible regions, then so does the quadratic objective.

It therefore remains to bound the difference between the linear maxima
uniformly over this family.
The difficulty is that the feasible region itself depends on the sample.
Our analysis bounds this difference uniformly by drawing on tools
from statistical learning theory like symmetrization and contraction.

We extend this approach to Max-$r$-Cut in Section~\ref{sec:main}.
For a bipartition, choosing one part completely determines the other.
But with more than two parts, the vertices not in one part
can be divided among the other $r-1$ parts in many ways,
and our analysis must track the parts and their joint interactions.
This makes the proof more involved, although its main stages remain the same.
We express the two (normalized) optima as evaluations of the same function
and then apply symmetrization and contraction to handle the random feasible set.

\subsection{Related work}
\label{sec:related}

\paragraph{Estimating the optimum.}
Goldreich, Goldwasser, and Ron~\cite{ggr98} gave the first
sublinear-time algorithm achieving an additive $\varepsilon n^2$-approximation
for unweighted Max-Cut using $O(\varepsilon^{-7}\log^2(2/\varepsilon))$ matrix queries.
Their combinatorial algorithm enumerates partitions of small vertex samples to define
candidate cuts, then estimates their values using additional random queries.
Their analysis also gives, as a corollary, the first value coreset
of size $\widetilde O(\varepsilon^{-5})$.

Alon, Fernandez de la Vega, Kannan, and Karpinski~\cite{afkk03}
improved the sample bound,
showing that the rescaled Max-Cut value of the induced subgraph
on $\widetilde O(\varepsilon^{-4})$ uniformly sampled vertices
gives an additive $\varepsilon n^2$-approximation. 
Their analysis combines
matrix decomposition into a small number of cut matrices,
due to Frieze and Kannan~\cite{fk99}, with linear programming.
Rudelson and Vershynin~\cite{rv07} removed the logarithmic factor using tools from
geometric functional analysis, showing that $O(\varepsilon^{-4})$ sampled vertices suffice.
Mathieu and Schudy~\cite{ms08} obtain the same sample bound for unweighted graphs
by analyzing a greedy algorithm.

This sampling approach was extended to non-dense graphs
by Bhaskara, Daruki, and Venkatasubramanian~\cite{bdv18}.
They show that degree-based importance sampling
in unweighted graphs of average degree $\bar d$ yields
a $(1\pm\varepsilon)$-approximate value coreset
with $O(n\log n/(\bar d\varepsilon^4))$ vertices in expectation.
Their analysis compares LP approximations of the original
and sampled graphs using LP duality and concentration bounds.

For unweighted Max-$r$-Cut, Goldreich, Goldwasser, and Ron~\cite{ggr98} obtained an
additive $\varepsilon n^2$-approximation using $O(\varepsilon^{-7}\log^2(2r/\varepsilon))$
matrix queries.
Andersson and Engebretsen~\cite{ae02} extended their combinatorial
sampling approach to finite-domain CSPs.
For Max-$r$-Cut, they obtain a comparable bound of
$O(\varepsilon^{-7}\log(2r)\log^2(2r/\varepsilon))$ queries.
Karpinski and Mark\'o~\cite{km16} used cut decompositions and LP duality
to analyze the rescaled optimum of
a uniformly sampled induced subgraph, i.e., a value coreset,
and showed that $\widetilde O(r^8\varepsilon^{-4})$ sampled vertices
suffice for the same additive guarantee.

\paragraph{Metric importance sampling.}
Fernandez de la Vega and Kenyon~\cite{vegakenyon01} obtained a
polynomial-time approximation scheme (PTAS) for metric Max-Cut
by reducing it to dense weighted Max-Cut.
Their reduction replicates each vertex according
to its weighted degree and rescales the edge weights.
Chen, Jiang, and Krauthgamer~\cite{cjk23} combined this reduction
with uniform-sampling bounds to show that importance sampling of
$O(\varepsilon^{-4})$ vertices
gives a
$(1\pm\varepsilon)$-approximate value coreset, and
they also implement this approach
in the streaming
setting for Euclidean Max-Cut.

\paragraph{Constructing a cut.}
The first PTASes for dense Max-Cut were obtained independently by
Arora, Karger, and Karpinski~\cite{akk99} and Fernandez de la Vega~\cite{vega96}.
Fernandez de la Vega uses a small vertex sample to guide the assignment 
of the remaining vertices, searching exhaustively over assignments within small blocks.
Arora, Karger, and Karpinski

enumerate assignments to $O(\varepsilon^{-2}\log n)$ sampled vertices,
use each assignment to construct an LP 
on the full graph, and round its solution.
Fotakis, Lampis, and Paschos~\cite{flp16}
extended this approach to unweighted graphs with
$\Omega(n^{1+\sigma})$ edges, for fixed $0<\sigma\le1$,
and achieved
approximation ratio $1-\varepsilon$ in time $2^{O(n^{1-\sigma}\log n/\varepsilon^3)}$.

Mathieu and Schudy~\cite{ms08} obtained additive
$O(\varepsilon n^2)$-approximation in expectation for unweighted graphs
in faster time $O(n^2)+2^{O(\varepsilon^{-2})}$.
They greedily extend every partition of a small random seed
within a larger sample, then extend the best resulting cut to the full graph.
Vertices are processed in random order and placed where they create more crossing
edges.
Yaroslavtsev~\cite{yar14} obtained the same approximation in faster time
$O(n\varepsilon^{-2})+2^{O(\varepsilon^{-2})}$,
by estimating the greedy gains using samples of already assigned vertices.

Yoshida and Zhou~\cite{yz14} use the Sherali--Adams hierarchy of LP relaxations
and show that
an optimal solution at level $O((\rho\varepsilon^2)^{-1})$
can be rounded to give a
$(1-\varepsilon)$-approximation on unweighted graphs with at least $\rho n^2$ edges,
for fixed $\rho>0$. For metric Max-Cut, $O(\varepsilon^{-2})$ levels suffice.

\paragraph{Cut oracles.}
A \emph{cut oracle} is an implicit representation (data structure)
of a cut of value at least $(1-\varepsilon)\mathrm{OPT}$
that reports, given a query vertex, its assignment (side) in the cut.
For dynamic streams of Euclidean points in $[\Delta]^d$, Lammersen, Sidiropoulos, and
Sohler~\cite{lss09} combine dimension reduction with the cut oracle of Frahling and
Sohler~\cite{fs05},
to obtain a $(1-\varepsilon)$-approximate cut oracle using
$\exp\!\bigl(\operatorname{poly}(\varepsilon^{-1})\bigr)\operatorname{poly}(d\log\Delta)$
space.
Menand and Waingarten~\cite{mw26} combine a sampled version of
the greedy algorithm of Mathieu and Schudy~\cite{ms08}
with the streaming sampler of Chen, Jiang, and Krauthgamer~\cite{cjk23},
to obtain such an oracle using the smaller
space bound $\poly(d\log\Delta/\varepsilon)$.
We do not provide a cut oracle.

\section{Sample Complexity for Max-Cut}
\label{sec:max-cut-sample-complexity}

In this section, we prove our main theorem that sampling \(O(\varepsilon^{-2})\) vertices achieves an additive \(\varepsilon n^2\)-approximation to the Max-Cut value.
We start by setting up the necessary notation
and state our main result in \Cref{thm:main}.
We then give an overview of the proof (in \Cref{sec:overview}),
which is followed by the proof itself:
We express the two cut values through a common functional (in \Cref{sec:cut-functional}),
then reduce the comparison of these two cut values
to a certain difference between two linear maximization problems (in \Cref{sec:cut-comparison}),
and then bound the expectation of this difference
(in \Cref{sec:sampling-error}).
Finally, we specialize the error bound
to uniform and importance sampling (in \Cref{sec:sampling-width}).

Let \(V\) be a vertex set of size $n$, and let
\(A\in\mathbb R^{V\times V}\) be a symmetric matrix with zero diagonal and nonnegative
edge weights, i.e., \(A_{uv}\ge0\) for all \(u,v\in V\).
Let \(d(v):=\sum_{u\in V}A_{vu}\) be the weighted degree of \(v\) and
let \(D:=\sum_{v\in V}d(v)\).
Isolated vertices contribute zero to every cut, so we may discard them and assume
\(d(v)>0\) for all \(v\in V\).

Fix an integer \(k\ge1\) and a positive probability vector \(p\in\mathbb R^V\).
Importance sampling allows \(p_v\ge\frac{d(v)}{\lambda D}\) for all
\(v\in V\), where \(\lambda\ge1\) is a parameter, and uniform sampling uses \(p_v=1/n\).
Draw \(X_1,\ldots,X_k\in V\) independently with replacement according to \(p\). If the
same vertex is sampled more than once, its copies remain distinct vertices of the
output.
The sampled graph has vertex set \([k]\), representing \((X_1,\ldots,X_k\)).
Define its edge weights \(\widehat A\) and a matrix \(K\in\mathbb R^{V\times V}\) by
\begin{equation}\label{eq:estimator}
  \widehat A_{ij}:=\frac{A_{X_iX_j}}{k^2p_{X_i}p_{X_j}},
  \qquad
  K_{uv}:=\frac{A_{uv}}{n^2p_up_v}.
\end{equation}

Let \(P_X\in\mathbb R^V\) be the empirical distribution of the sample,
that is \(P_X(v):=|\{i:X_i=v\}|/k\) for every \(v\in V\).
This matrix \(K\) makes it convenient to express and compare the Max-Cut values in the
original and sampled graphs, after scaling by a factor of \(1/n^2\). The expressions
below differ only in whether they use the sampling distribution \(p\) or its empirical
distribution \(P_X\).
\[
  \frac{\MC(A)}{n^2}
  =\max_{S\subseteq V}\sum_{u\in S}\sum_{v\notin S}p_up_vK_{uv},
  \qquad
  \frac{\MC(\widehat A)}{n^2}
  =\max_{S\subseteq V}\sum_{u\in S}\sum_{v\notin S}P_X(u)P_X(v)K_{uv}.
\]
The second identity uses the fact that in the sampled graph, at least one optimal cut
keeps all copies of the same vertex together. Lemma~\ref{lem:value-identities} proves
this fact and both identities.

We will use the following quantity to control the sampling error through second moment
estimates.
\[
  H:=\sum_{v\in V}p_v
  \left(\sum_{u\in V}p_uK_{uv}^2\right)^{1/2}.
\]
A simple calculation (see Lemma~\ref{lem:degree-width}) shows that \(H\le1\) whenever
every edge weight \(A_{uv}\) lies in \([0,1]\) and the sampling distribution satisfies
\begin{equation}\label{eq:probability-condition}
  p_v\ge\frac{d(v)}{n^2},
  \qquad\forall\,v\in V.
\end{equation}

\begin{theorem}[Sampling approximates the Max-Cut value]\label{thm:main}
For every positive probability vector \(p\in\mathbb R^V\),
\begin{equation}\label{eq:main-expectation}
  \E\left|
    \frac{\MC(\widehat A)-\MC(A)}{n^2}
  \right|
  \le \frac{48H}{\sqrt{k}}.
\end{equation}
\end{theorem}

\begin{corollary}[High-probability bound]\label{cor:tail}
Assume \(0\le A_{uv}\le1\) for every \(u,v\in V\). Then for every positive sampling
distribution satisfying \eqref{eq:probability-condition}, including uniform and
importance sampling with \(\lambda=1\),
\begin{equation}\label{eq:tail-main}
  \forall\,0<\varepsilon<1,\quad
  \Prb\left(
    |\MC(\widehat A)-\MC(A)|\ge\varepsilon n^2
  \right)
  \le\frac{48}{\varepsilon\sqrt{k}}.
\end{equation}
Consequently, to get error probability at most \(\delta\in(0,1)\), it suffices to take
any \(k\ge2304/(\delta^2\varepsilon^2)\).
\end{corollary}

\begin{proof}
\Cref{sec:sampling-width} verifies the bound \(H\le1\) for sampling that satisfies
\eqref{eq:probability-condition}, and then Markov's inequality implies
(\ref*{eq:tail-main}). Solving for \(k\) gives the sample size bound for given
\(\delta\).
\end{proof}

A simple calculation shows that, for importance sampling with parameter \(\lambda\ge1\),
an extra factor of \(\lambda\) in the sample size suffices, e.g.,
\(k\ge20736\lambda\varepsilon^{-2}\) gives error probability at most \(1/3\).

\subsection{Proof overview}
\label{sec:overview}

Our first step is to view sampling as a change in vertex weights. Given \(K\) and a
probability vector \(r\in\mathbb R^V\), we define \(F_K(r)\) as the maximum cut value
with respect to \(K\) and vertex weights \(r_v\), meaning that each edge \((u,v)\) has
weight \(r_ur_vK_{uv}\). It will be convenient to allow fractional cuts, where each
vertex's weight \(r_v\) may be split between the two sides, although we show in
\cref{sec:cut-functional} that this does not change the optimum value, and in fact
\[
  F_K(p)=\frac{\MC(A)}{n^2},
  \qquad
  F_K(P_X)=\frac{\MC(\widehat A)}{n^2}.
\]

We then compare in \cref{sec:cut-comparison} these two values, and more generally
\(F_K(r)\) and \(F_K(s)\) for any two probability vectors \(r,s\in\mathbb R^V\). We do
this by matching fractional cuts of the two vertex weights \(r\) and \(s\). For either
\(q\in\{r,s\}\), a fractional cut is represented by \(0\le a\le q\) and has value
\(a^\top K(q-a)\). We relax \(a^\top K(q-a)\) to \(a^\top Kc\), allowing
\(c\in\mathbb R^V\) to vary independently of \(a\) with \(0\le c\le r\) or
\(0\le c\le s\). For fixed \(c\), maximizing over \(a\) gives
\(M_q(c):=\max_{0\le a\le q}a^\top Kc\), and our analysis compares \(M_r(c)\) and
\(M_s(c)\) for the same \(c\). Let \(\Delta_K(r,s)\) be their largest absolute
difference over a suitable region \(\mathcal C(r,s)\) of vectors \(c\), i.e.,
\[
  \Delta_K(r,s)
  :=\max_{c\in\mathcal C(r,s)}\bigl|M_r(c)-M_s(c)\bigr|.
\]

A minimax argument converts the comparison of these maxima into a matching of fractional
cuts. For each \(a\in\mathbb R^V\) with \(0\le a\le r\), it produces a matching
\(b\in\mathbb R^V\), \(0\le b\le s\), such that \(a^\top Kc\) and \(b^\top Kc\) are
close for all feasible \(c\) at once, meaning that
\[
  \max_c
  \bigl|(a-b)^\top Kc\bigr|
  \le\Delta_K(r,s).
\]
Using this matching, a simple calculation shows that
\[
  |F_K(r)-F_K(s)|\le4\Delta_K(r,s).
\]

Applying this bound with \(r=p\) and \(s=P_X\) reduces the problem to bounding
\(\E_X\Delta_K(p,P_X)\), which we address in \cref{sec:sampling-error}. The main
difficulty is that both the sampling error for a given
\(c\) and the feasible region for \(c\) depend on the sample \(X\). The maximizing
vector can therefore adapt to the sample, so a bound for each \(c\) fixed in advance
does not directly control \(\Delta_K(p,P_X)\). Similar difficulties arise in statistical
learning when comparing a hypothesis's error on the sample used to choose it with its
error under the underlying distribution. We borrow from that domain a standard tool
called \emph{symmetrization}. We draw an independent \emph{ghost sample}
\(Y_1,\ldots,Y_k\) from \(p\), solely for the sake of analysis, and denote its empirical
distribution by \(P_Y\) and use it to rewrite \(\Delta_K(p,P_X)\) as
\[
  \Delta_K(p,P_X)
  =
  \max_{c\in\mathcal C(p,P_X)}
  \left|\E_Y\bigl[M_{P_X}(c)-M_{P_Y}(c)\bigr]\right|.
\]

We enlarge the feasible region for \(c\) to one that is invariant under swaps between
\(X_i\) and \(Y_i\), and then draw \(k\) independent random signs
\(\sigma_1,\ldots,\sigma_k\in\{\pm1\}\), where \(+1\) leaves \(X_i,Y_i\) in place, and
\(-1\) exchanges them. Crucially, the enlarged region does not change under these
exchanges. Writing \([t]_+:=\max\{t,0\}\), the swaps turn \(M_{P_X}(c)-M_{P_Y}(c)\) into
\[
  \frac1k\sum_{i=1}^k\sigma_i
    \bigl([(Kc)_{X_i}]_+-[(Kc)_{Y_i}]_+\bigr).
\]
This lets us apply Ledoux--Talagrand contraction to reduce to a signed sum linear in
\(c\). We can therefore remove the maximum over \(c\) by bounding each
\(\lvert c_v\rvert\) by its worst-case value. A simple second moment calculation then
gives
\[
  \E_X\Delta_K(p,P_X)\le\frac{12H}{\sqrt{k}}.
\]

\subsection{Reduce both Max-Cut values to one functional}
\label{sec:cut-functional}

We express the normalized Max-Cut values of the original graph \(A\) and the sampled
graph \(\widehat A\) as evaluations of the same functional \(F_K(\cdot)\) at \(p\) and
\(P_X\), respectively.

To define \(F_K\), we first describe fractional cuts. Cuts are usually represented by
\(x_v\in\{0,1\}\), where vertices \(v\) with \(x_v=1\) are on one side and those with
\(x_v=0\) on the other side. Allowing \(x_v\in[0,1]\) gives a fractional cut, in which a
fraction \(x_v\) of vertex \(v\) goes to one side and the remaining fraction \(1-x_v\)
to the other side. With vertex weights \(r_v\ge0\), the weight of \(v\) placed on one
side is \(a_v=r_vx_v\), which we call the \emph{weighted fraction} of \(v\) on that
side. For simplicity, we will simply refer to \(a_v\) as the fraction of \(v\) on that
side. The other side receives weight \(r_v-a_v\). Thus a fractional cut has
\(0\le a_v\le r_v\), and an ordinary cut has \(a_v\in\{0,r_v\}\). The probability
vectors \(p\) and \(P_X\) provide the vertex weights for the original and sampled graphs
respectively. Using fractional cuts in a vertex-weighted graph lets us express both cut
values through the same functional.

Define the feasible fractional cuts by
\[
  Z(r):=\{a\in\mathbb R^V:0\le a_v\le r_v\text{ for every }v\}
\]

and their optimal value by
\begin{equation}\label{eq:F-definition}
  F_K(r):=\max_{a\in Z(r)}a^\top K(r-a).
\end{equation}
Allowing fractional cuts makes the feasible set convex, which is useful for the
comparison in \cref{sec:cut-comparison}.

\begin{lemma}[The maximum is attained by a cut]\label{lem:corner-maximizer}
For every nonnegative weight vector \(r\in\mathbb R^V\), there exists a maximizer
\(a\in Z(r)\) in \eqref{eq:F-definition} that corresponds to a cut, i.e.,
\(a_v\in\{0,r_v\}\) for all \(v\in V\).
\end{lemma}

\begin{proof}
Since \(K_{vv}=0\), the objective \(a^\top K(r-a)\) is affine in each variable \(a_v\)
(when the other variables are fixed). An affine function on \([0,r_v]\) attains its
maximum at an endpoint, so we can move \(a_v\) to an endpoint without decreasing the
objective. Starting from a maximizer and applying this operation iteratively yields a
maximizer that has \(a_v\in\{0,r_v\}\) for every \(v\).
\end{proof}

\begin{lemma}[The two Max-Cut values]\label{lem:value-identities}
With \(p\), \(P_X\), and \(K\) defined above,
\[
  F_K(p)=\frac{\MC(A)}{n^2},
  \qquad
  F_K(P_X)=\frac{\MC(\widehat A)}{n^2}.
\]
\end{lemma}

\begin{proof}
For the first identity, Lemma~\ref{lem:corner-maximizer} lets us maximize over cuts
\(S\subseteq V\). Since \(p_up_vK_{uv}=A_{uv}/n^2\),
\[
  F_K(p)
  =
  \max_{S\subseteq V}\sum_{u\in S}\sum_{v\notin S}p_up_vK_{uv}
  =
  \frac{\MC(A)}{n^2}.
\]

For the second identity, we first show that all copies of a repeatedly sampled vertex
can be moved to the same side without decreasing the cut value. Start with an optimal
cut for the sampled subgraph (induced on \(X_1,\ldots,X_k\)).

For an original vertex \(v\), its sampled copies have no edges between them and have the
same edge weights to all other sampled vertices. Holding those other vertices fixed,
choose a side maximizing the cut contribution of one copy of \(v\). The same choice
maximizes the contribution of every copy. If the two sides tie, choose either side
arbitrarily. Put all copies on that side and repeat for each original vertex. These
moves cannot decrease the cut value, so the final cut is still optimal and keeps all
copies of each vertex together.

Such a cut is described by a set \(S\subseteq V\). All copies of vertices in \(S\) lie
on one side, and all remaining copies lie on the other side. For \(u\in S\) and
\(v\notin S\), there are \(kP_X(u)\,kP_X(v)\) crossing pairs of copies. Each pair
\((i,j)\) with \(X_i=u\) and \(X_j=v\) contributes to the cut value divided by \(n^2\)
as follows:
\[
  \frac{\widehat A_{ij}}{n^2}
  =\frac{A_{uv}}{k^2n^2p_up_v}
  =\frac{K_{uv}}{k^2}.
\]
  Therefore, applying Lemma~\ref{lem:corner-maximizer} with \(r=P_X\),
\[
  F_K(P_X)
  =\max_{S\subseteq V}
    \sum_{u\in S}\sum_{v\notin S}P_X(u)P_X(v)K_{uv}
  =\frac{\MC(\widehat A)}{n^2}.
\]
\end{proof}

By Lemma~\ref{lem:value-identities},
\[
  \frac{|\MC(\widehat A)-\MC(A)|}{n^2}
  =
  \left|F_K(P_X)-F_K(p)\right|.
\]

\subsection{Compare cut values under two vertex distributions}
\label{sec:cut-comparison}

In this subsection, we bound the difference between \(F_K(r)\) and \(F_K(s)\) for two
probability vectors \(r,s\) on \(V\). In our intended application, we will use \(r=p\)
and \(s=P_X\).

For a nonnegative vector \(q\), let \(M_q(c)\) be the maximum of \(x^\top Kc\) over
\(x\in Z(q)\), namely
\begin{equation}\label{eq:profile}
  M_q(c)
  :=\max_{x\in Z(q)}x^\top Kc
  =\sum_vq_v[(Kc)_v]_+,
  \qquad \text{where }[t]_+:=\max\{t,0\}.
\end{equation}
The equality follows by choosing \(x_v=q_v\) when \((Kc)_v>0\) and \(x_v=0\) otherwise.

To compare \(M_r\) and \(M_s\), we use a larger domain that contains both \(Z(r)\) and
\(Z(s)\) and is symmetric and convex.
\begin{equation}\label{eq:Crs}
  \mathcal C(r,s)
  :=
  \operatorname{conv}\bigl(
    Z(r)\cup-Z(r)\cup Z(s)\cup-Z(s)
  \bigr).
\end{equation}
We consider their largest absolute difference, defined by
\begin{equation}\label{eq:delta}
  \Delta_K(r,s)
  :=
  \max_{c\in\mathcal C(r,s)}
  |M_r(c)-M_s(c)|.
\end{equation}

\begin{lemma}[Comparison of cut values]
\label{lem:value-stability}
For every pair of probability vectors \(r,s\in\mathbb R^V\),
\begin{equation}\label{eq:value-stability}
  |F_K(r)-F_K(s)|\le4\Delta_K(r,s).
\end{equation}
\end{lemma}

Taking \(r=p\) and \(s=P_X\), this lemma gives \(|F_K(P_X)-F_K(p)|\le4\Delta_K(p,P_X)\).
It then remains to bound \(\E_X\Delta_K(p,P_X)\), which we do in
\cref{sec:sampling-error}.

To prove Lemma~\ref{lem:value-stability}, we first prove two bounds. One matches
fractional cuts under \(r\) and \(s\), and the other compares the vertex masses.

\begin{lemma}[Matching fractional cuts]
\label{lem:profile-matching}
Let \(r,s\in\mathbb R^V\) be probability vectors. For every \(a\in Z(r)\), there exists
\(b\in Z(s)\) such that
\begin{equation}\label{eq:match-a-b}
  \max_{c\in\mathcal C(r,s)}
  |(a-b)^\top Kc|
  \le\Delta_K(r,s).
\end{equation}

Consequently,
\begin{equation}\label{eq:mass-match}
  \max_{c\in\mathcal C(r,s)}
  |(r-s)^\top Kc|
  \le2\Delta_K(r,s).
\end{equation}
\end{lemma}

\begin{proof}
Fix \(a\in Z(r)\). Using \(\mathcal C(r,s)=-\mathcal C(r,s)\) and applying Sion's
minimax theorem \cite{sion} to the function \((b,c)\mapsto(a-b)^\top Kc\), which is
affine in each variable, on the compact convex sets \(Z(s)\) and \(\mathcal C(r,s)\), we
obtain
\begin{align*}
  \min_{b\in Z(s)}
  \max_{c\in\mathcal C(r,s)}|(a-b)^\top Kc|
  &=
  \min_{b\in Z(s)}
  \max_{c\in\mathcal C(r,s)}(a-b)^\top Kc\\
  &=
  \max_{c\in\mathcal C(r,s)}
  \min_{b\in Z(s)}(a-b)^\top Kc\\
  &=
  \max_{c\in\mathcal C(r,s)}
  \left(a^\top Kc-M_s(c)\right)\\
  &\le
  \max_{c\in\mathcal C(r,s)}
  \left(M_r(c)-M_s(c)\right)
  \le\Delta_K(r,s).
\end{align*}

The minimum is attained, so \eqref{eq:match-a-b} follows.

To prove the second inequality, fix \(c\in\mathcal C(r,s)\). The identity
\([t]_+-[-t]_+=t\) and \eqref{eq:profile} give
\[
  (r-s)^\top Kc
  =
  \bigl(M_r(c)-M_s(c)\bigr)
  -
  \bigl(M_r(-c)-M_s(-c)\bigr).
\]
Both \(c\) and \(-c\) belong to \(\mathcal C(r,s)\), so each of the two differences on
the right has absolute value at most \(\Delta_K(r,s)\).

The triangle inequality gives \(\lvert(r-s)^\top Kc\rvert\le2\Delta_K(r,s)\), proving
\eqref{eq:mass-match}.
\end{proof}

\begin{proof}[Proof of Lemma~\ref{lem:value-stability}]
Let \(a\in Z(r)\) attain \(F_K(r)\), and choose \(b\in Z(s)\) as in
Lemma~\ref{lem:profile-matching}. Write \(\Delta=\Delta_K(r,s)\).

For every \(c\in\mathcal C(r,s)\), Lemma~\ref{lem:profile-matching} gives
\begin{equation}\label{eq:cut-comparison-terms}
  |(a-b)^\top Kc|\le\Delta,
  \qquad
  |(r-s)^\top Kc|\le2\Delta.
\end{equation}
Since \(F_K(s)\ge b^\top K(s-b)\), the symmetry of \(K\) gives
\begin{equation}\label{eq:cut-value-expansion}
  \begin{aligned}
  F_K(r)-F_K(s)
  &\le a^\top K(r-a)-b^\top K(s-b)\\
  &=(a-b)^\top K(r-a)+(r-s)^\top Kb-(a-b)^\top Kb\\
  &\le |(a-b)^\top K(r-a)|
      +|(r-s)^\top Kb|+|(a-b)^\top Kb|.
  \end{aligned}
\end{equation}
Since \(r-a\in Z(r)\) and \(b\in Z(s)\), both lie in \(\mathcal C(r,s)\), and we can
apply \eqref{eq:cut-comparison-terms} to get
\[
  F_K(r)-F_K(s)\le\Delta+2\Delta+\Delta=4\Delta.
\]
Interchanging \(r\) and \(s\) gives the same bound for \(F_K(s)-F_K(r)\), proving
\eqref{eq:value-stability}.
\end{proof}

\subsection{Bound the sampling error}
\label{sec:sampling-error}

In this subsection we prove that
\begin{equation}\label{eq:section4-goal}
  \E_X\Delta_K(p,P_X)
  \le
  \frac{12H}{\sqrt{k}}.
\end{equation}
Together with Lemmas ~\ref{lem:value-identities} and \ref{lem:value-stability}, this
completes the proof of Theorem~\ref{thm:main}. Recall that \(\Delta_K(p,P_X)\) is the
largest absolute difference between the exact and sample averages
\[
  M_p(c)
  =
  \sum_vp_v[(Kc)_v]_+,
  \qquad
  M_{P_X}(c)
  =
  \frac1k\sum_{i=1}^k[(Kc)_{X_i}]_+,
\]
over \(c\in\mathcal C(p,P_X)\). The challenge is that this domain depends on the same
sample \(X\) used in \(M_{P_X}\).

The proof has two steps, using several tools from statistical learning theory.

\begin{enumerate}
  \item \textbf{Symmetrization.}
  Instead of comparing \(P_X\) with its expectation \(p\), we compare the empirical
  distributions \(P_X,P_Y\) of two independent samples \(X,Y\). We enlarge the domain of
  \(c\) to a box \(\mathcal B_{X,Y}\) \cite{mendelson-philips}.

  We use symmetrization and Ledoux--Talagrand contraction \cite{ledoux-talagrand} to
  reduce the problem to bounding the expected maximum of sums of \((Kc)_{X_i}\) with
  independent random signs\nobreakspace(Lemma~\ref{lem:adaptive-symmetrization}).

  \item \textbf{Second-moment bounds.}
  The signed sums are linear in \(c\), so we bound their maximum over the box one
  coordinate at a time. We bound the maximum by three sums, weighted by \(p\), \(P_X\),
  and \(P_Y\). We bound each sum's expectation by \(H/\sqrt{k}\) using the upper \(L^1\)
  Khintchine bound, obtaining a total of at most \(3H/\sqrt{k}\)
  (Lemma~\ref{lem:pooled-box-width}).
\end{enumerate}
Combining the two estimates gives \eqref{eq:section4-goal}.
Lemma~\ref{lem:value-stability} then bounds the expected normalized Max-Cut error by
\(48H/\sqrt{k}\).

\subsection*{Step 1. Symmetrization}

\begin{lemma}[Reduction to a signed linear maximum]
\label{lem:adaptive-symmetrization}
Let \(Y_1,\ldots,Y_k\) be independent draws according to \(p\), independent of \(X\),
with sample-frequency vector \(P_Y\). Let \(\sigma_1,\ldots,\sigma_k\in\{\pm1\}\) be
independent uniform signs, independent of \(X,Y\), and define
\[
  \mathcal B_{X,Y}
  :=
  \left\{
    c\in\mathbb R^V:
    \forall v\in V,\
    |c_v|\le p_v+P_X(v)+P_Y(v)
  \right\}.
\]
Then
\begin{equation}\label{eq:delta-to-linear}
  \E_X\Delta_K(p,P_X)
  \le
  4\E_{X,Y,\sigma}
  \max_{c\in\mathcal B_{X,Y}}
  \left|
    \frac1k\sum_{i=1}^k\sigma_i(Kc)_{X_i}
  \right|.
\end{equation}
\end{lemma}

\begin{proof}
We begin with the symmetrization step using a ghost sample. For every
\(c\in\mathcal C(p,P_X)\), the independent sample average has mean \(M_p(c)\), since
\[
  \E_YM_{P_Y}(c)
  =
  \sum_vp_v[(Kc)_v]_+
  =
  M_p(c).
\]
The box \(\mathcal B_{X,Y}\) is convex and contains each of the four sets defining
\(\mathcal C(p,P_X)\), so
\begin{equation}\label{eq:C-in-B}
  \mathcal C(p,P_X)\subseteq\mathcal B_{X,Y}.
\end{equation}
With \(X\) fixed, \(\mathcal C(p,P_X)\) does not depend on \(Y\), the absolute value of
an expectation is at most the expectation of the absolute value and the maximum of
expectations is at most the expectation of the maximum (Jensen's inequality). Thus
\begin{align*}
  \Delta_K(p,P_X)
  &=
  \max_{c\in\mathcal C(p,P_X)}
  \left|
    \E_Y\bigl[M_{P_X}(c)-M_{P_Y}(c)\bigr]
  \right|\\
  &\le
  \max_{c\in\mathcal C(p,P_X)}
  \E_Y|M_{P_X}(c)-M_{P_Y}(c)|\\
  &\le
  \E_Y
  \max_{c\in\mathcal C(p,P_X)}
  |M_{P_X}(c)-M_{P_Y}(c)|\\
  &\le
  \E_Y
  \max_{c\in\mathcal B_{X,Y}}
  \left|
    \frac1k\sum_{i=1}^k
    \left(
      [(Kc)_{X_i}]_+-[(Kc)_{Y_i}]_+
    \right)
  \right|.
\end{align*}

We next take expectation over \(X\) and use pairwise exchanges to express the expected
maximum in terms of sums with random signs (Rademacher symmetrization). For each fixed
sign vector \(\sigma\in\{\pm1\}^k\), set
\[
  (\widetilde X_i,\widetilde Y_i)
  :=
  \begin{cases}
    (X_i,Y_i),&\sigma_i=1,\\
    (Y_i,X_i),&\sigma_i=-1.
  \end{cases}
\]
Each exchange reverses its pair's contribution, so
\[
  [(Kc)_{\widetilde X_i}]_+-[(Kc)_{\widetilde Y_i}]_+
  =
  \sigma_i\bigl([(Kc)_{X_i}]_+-[(Kc)_{Y_i}]_+\bigr).
\]
The exchanges preserve the joint distribution of the samples. They also preserve
\(P_X+P_Y\), and thus \(\mathcal B_{\widetilde X,\widetilde Y}=\mathcal B_{X,Y}\).

Together with the preceding bound, these identities give
\begin{align}
  \E_X\Delta_K(p,P_X)
  &\le
  \E_{X,Y}
  \max_{c\in\mathcal B_{X,Y}}
  \left|
    \frac1k\sum_{i=1}^k
    \bigl([(Kc)_{X_i}]_+-[(Kc)_{Y_i}]_+\bigr)
  \right|\notag\\
  &=
  \E_{X,Y,\sigma}
  \max_{c\in\mathcal B_{X,Y}}
  \left|
    \frac1k\sum_{i=1}^k\sigma_i
    \left(
      [(Kc)_{X_i}]_+-[(Kc)_{Y_i}]_+
    \right)
  \right|\notag\\
  &\le
  2\E_{X,Y,\sigma}
  \max_{c\in\mathcal B_{X,Y}}
  \left|
    \frac1k\sum_{i=1}^k\sigma_i[(Kc)_{X_i}]_+
  \right|,\label{eq:sample-symmetrization}
\end{align}
where the last inequality uses the triangle inequality and symmetry between \(X\) and
\(Y\). Finally, \(g:t\mapsto[t]_+\) is \(1\)-Lipschitz and vanishes at \(0\).
For fixed \(X,Y\), the Ledoux--Talagrand contraction inequality
(Lemma~\ref{lem:absolute-rademacher-contraction}; see \cite{ledoux-talagrand}) therefore gives
\[
  \E_\sigma
  \max_{c\in\mathcal B_{X,Y}}
  \left|
    \frac1k\sum_{i=1}^k\sigma_i[(Kc)_{X_i}]_+
  \right|
  \le
  2\E_\sigma
  \max_{c\in\mathcal B_{X,Y}}
  \left|
    \frac1k\sum_{i=1}^k\sigma_i(Kc)_{X_i}
  \right|.
\]
Taking expectation over \(X,Y\) and substituting into \eqref{eq:sample-symmetrization}
proves the lemma.
\end{proof}

\subsection*{Step 2. Second-moment bounds}

\begin{lemma}[Bound for the signed maximum]
\label{lem:pooled-box-width}
\begin{equation}\label{eq:pooled-box-width}
  \E_{X,Y,\sigma}
  \max_{c\in\mathcal B_{X,Y}}
  \left|
    \frac1k\sum_{i=1}^k\sigma_i(Kc)_{X_i}
  \right|
  \le\frac{3H}{\sqrt{k}}.
\end{equation}
\end{lemma}

\begin{proof}
We first bound the maximum. Fix \(X,Y,\sigma\), and let \(b_v\) denote the bound on the
\(v\)-th coordinate in the definition of \(\mathcal B_{X,Y}\), so
\[
  b_v:=p_v+P_X(v)+P_Y(v).
\]
Since \(K\) is symmetric, every \(c\in\mathcal B_{X,Y}\) satisfies
\begin{equation}\label{eq:signed-average-as-matrix}
  \left|\sum_{i=1}^k\sigma_i(Kc)_{X_i}\right|
  =\left|\sum_v c_v\sum_i\sigma_iK_{vX_i}\right|
  \le\sum_vb_v\left|\sum_i\sigma_iK_{vX_i}\right|.
\end{equation}

Taking the maximum over \(c\in\mathcal B_{X,Y}\) and substituting the definition of
\(b_v\), we obtain
\begin{equation}\label{eq:three-weight-split}
  \max_{c\in\mathcal B_{X,Y}}
  \left|\sum_{i=1}^k\sigma_i(Kc)_{X_i}\right|
  \le
  \sum_vp_v\left|\sum_i\sigma_iK_{vX_i}\right|
  +\sum_vP_X(v)\left|\sum_i\sigma_iK_{vX_i}\right|
  +\sum_vP_Y(v)\left|\sum_i\sigma_iK_{vX_i}\right|.
\end{equation}
We have bounded the maximum over all \(c\) by three explicit sums. We bound them
separately because the weights from \(P_X\) depend on the signed sums, whereas the
weights from \(P_Y\) are independent of them.

\paragraph[Bound the sum weighted by p.]{Bound the sum weighted by \(p\).}
Since \(K\) is symmetric, we can write \(H=\sum_vp_vh_v\), where, for each \(v\in V\),
\[
  h_v:=\left(\sum_up_uK_{vu}^2\right)^{1/2}.
\]

For fixed \(X\), the upper \(L^1\) Khintchine inequality
(Lemma~\ref{lem:upper-l1-khintchine}) gives
\[
  \E_\sigma\left|\sum_i\sigma_iK_{vX_i}\right|
  \le
  \left(\sum_iK_{vX_i}^2\right)^{1/2}.
\]
Taking expectation over \(X\) and applying Jensen's inequality gives
\[
  \E_{X,\sigma}\left|\sum_i\sigma_iK_{vX_i}\right|
  \le
  \E_X\left(\sum_iK_{vX_i}^2\right)^{1/2}
  \le
  \left(\sum_i\E_XK_{vX_i}^2\right)^{1/2}
  =\sqrt{k}\,h_v,
\]

where the last equality uses \(\E_XK_{vX_i}^2=\sum_up_uK_{vu}^2=h_v^2\), since each
\(X_i\) is drawn according to \(p\).

We thus obtain that
\begin{equation}\label{eq:p-weighted-row}
  \E_{X,\sigma}\sum_vp_v\left|\sum_i\sigma_iK_{vX_i}\right|
  \le\sqrt{k}\sum_vp_vh_v=\sqrt{k}\,H.
\end{equation}

\paragraph[Bound the sum weighted by P_Y.]{Bound the sum weighted by \(P_Y\).}
We can use the preceding estimate directly for \(P_Y\), because the sample \(Y\) is
independent of \(X\) and \(\sigma\). Since \(\E_YP_Y(v)=p_v\), equation
\eqref{eq:p-weighted-row} gives
\begin{equation}\label{eq:PY-weighted-row}
  \E_{X,Y,\sigma}\sum_vP_Y(v)\left|\sum_i\sigma_iK_{vX_i}\right|
  =
  \E_{X,\sigma}\sum_vp_v\left|\sum_i\sigma_iK_{vX_i}\right|
  \le\sqrt{k}\,H.
\end{equation}

\paragraph[Bound the sum weighted by P_X.]{Bound the sum weighted by \(P_X\).}
We can write the \(P_X\)-weighted sum as
\[
  \sum_vP_X(v)\left|\sum_i\sigma_iK_{vX_i}\right|
  =
  \frac1k\sum_{\ell=1}^k\left|\sum_i\sigma_iK_{X_\ell X_i}\right|.
\]

Fix \(\ell\) and condition on \(X_\ell=u\). Since \(K_{uu}=0\), we have
\(\sum_i\sigma_iK_{X_\ell X_i}=\sum_{i\ne\ell}\sigma_iK_{uX_i}\). The other \(X_i\) are
independent draws according to \(p\), so applying Khintchine and Jensen again gives
\[
  \begin{aligned}
  \E_{X,\sigma}\left[
    \left|\sum_i\sigma_iK_{X_\ell X_i}\right|
    \mathrel{\Big|}X_\ell=u
  \right]
  &\le
  \E_X\left[
    \left(\sum_{i\ne\ell}K_{uX_i}^2\right)^{1/2}
    \mathrel{\Big|}X_\ell=u
  \right]\\
  &\le
  \left((k-1)h_u^2\right)^{1/2}
  =\sqrt{k-1}\,h_u.
  \end{aligned}
\]
We remove the conditioning on \(X_\ell=u\) by the law of total expectation. Summing the
resulting bounds over \(\ell\), dividing by \(k\), and using \(H=\sum_up_uh_u\), we
obtain
\begin{equation}\label{eq:PX-weighted-row}
  \E_{X,\sigma}\sum_vP_X(v)\left|\sum_i\sigma_iK_{vX_i}\right|
  \le
  \sqrt{k-1}\sum_up_uh_u=\sqrt{k-1}\,H
  \le
  \sqrt{k}\,H.
\end{equation}

Each of the three unnormalized sums has expected value at most \(\sqrt{k}\,H\).
Substituting these bounds into \eqref{eq:three-weight-split} and dividing by \(k\) gives
\(3H/\sqrt{k}\), proving Lemma~\ref{lem:pooled-box-width}.
\end{proof}
\subsection*{Putting it together}

\begin{theorem}[Expected error in \(F_K\)]\label{thm:sampling-value}
Let \(X_1,\ldots,X_k\) be independent draws from \(p\), and let \(P_X\) be their
empirical distribution. Then
\begin{equation}\label{eq:sampling-value}
  \E_X|F_K(P_X)-F_K(p)|
  \le\frac{48H}{\sqrt{k}}.
\end{equation}
\end{theorem}

\begin{proof}
Taking expectations in Lemma~\ref{lem:value-stability} with \(r=p\) and \(s=P_X\), and
using Lemmas ~\ref{lem:adaptive-symmetrization} and \ref{lem:pooled-box-width}, gives
\[
  \E_X|F_K(P_X)-F_K(p)|
  \le4\E_X\Delta_K(p,P_X)
  \le4\cdot4\cdot\frac{3H}{\sqrt{k}}
  =\frac{48H}{\sqrt{k}}.
\]
\end{proof}

\subsection[Bound H through the sampling probabilities]{Bound \(H\) through the sampling probabilities}
\label{sec:sampling-width}

The sampling probabilities determine \(H\) through the ratios \(d(v)/p_v\). We first
bound \(H\) in terms of these ratios, then apply the bound to importance and uniform
sampling.

\begin{lemma}[Bounding \(H\)]\label{lem:degree-width}
Let \(p\in\mathbb R^V\) be a positive probability vector, and define \(K\) and \(H\) as
above. If \(0\le A_{uv}\le1\) for all \(u,v\in V\), then
\begin{equation}\label{eq:degree-width}
  H^2\le\frac1{n^3}\sum_{v\in V}\frac{d(v)}{p_v}.
\end{equation}
In particular, if \(p_v\ge d(v)/n^2\) for every \(v\in V\), then \(H\le1\).
\end{lemma}

\begin{proof}
Substituting the definition of \(K\) into \(H\) cancels the probabilities outside the
square roots, giving
\begin{equation}\label{eq:h-expanded}
  H
  =
  \frac1{n^2}\sum_{v\in V}
  \left(\sum_{u\in V}\frac{A_{uv}^2}{p_u}\right)^{1/2}.
\end{equation}

Jensen's inequality and \(A_{uv}^2\le A_{uv}\) give
\[
  H^2
  \le\frac1{n^3}\sum_{u,v\in V}\frac{A_{uv}^2}{p_u}
  \le\frac1{n^3}\sum_{u\in V}\frac{d(u)}{p_u}.
\]
If \(p_v\ge d(v)/n^2\) for all \(v\in V\), then \(H^2\le n^{-3}\cdot n\cdot n^2=1\).
\end{proof}

Uniform and exact importance sampling both satisfy \(p_v\ge d(v)/n^2\). For exact
importance sampling, \(p_v=d(v)/D\) and \(D\le n^2\). For uniform sampling, \(p_v=1/n\)
and \(d(v)\le n\). Thus Lemma~\ref{lem:degree-width} gives \(H\le1\).

Keeping the dependence on \(D\) in \cref{lem:degree-width} gives a relative-error bound
in terms of the average weighted degree.

\begin{corollary}[Relative error from average weighted degree]
\label{cor:average-degree-relative-error}
Assume \(0\le A_{uv}\le1\) for all \(u,v\in V\). Let \(\bar d=D/n>0\) be the average
weighted degree, and write \(\mathrm{OPT}=\MC(A)\). Suppose the positive sampling
distribution \(p\) satisfies
\[
  \sum_{v\in V}\frac{d(v)}{p_v}\le\lambda nD
\]
for some \(\lambda\ge1\). Then
\[
  \E\left|\MC(\widehat A)-\mathrm{OPT}\right|
  \le192\sqrt{\frac{\lambda n}{\bar d k}}\,\mathrm{OPT}.
\]
\end{corollary}

\begin{proof}
We first bound \(H\) in terms of the average weighted degree. By \cref{lem:degree-width}
and the assumption on \(p\),
\[
  H^2
  \le\frac1{n^3}\sum_{v\in V}\frac{d(v)}{p_v}
  \le\frac{\lambda D}{n^2}
  =\frac{\lambda\bar d}{n}.
\]

A uniformly random cut has expected value \(D/4\), so \(\mathrm{OPT}\ge n\bar d/4\).
Substituting these two bounds into \cref{thm:main} gives
\[
  \frac{\E|\MC(\widehat A)-\mathrm{OPT}|}{\mathrm{OPT}}
  \le\frac{48n^2H}{\mathrm{OPT}\sqrt{k}}
  \le192\sqrt{\frac{\lambda n}{\bar d k}}.
\]
\end{proof}

Importance sampling with parameter \(\lambda\) satisfies the corollary's assumption,
since \(p_v\ge d(v)/(\lambda D)\) gives \(\sum_v d(v)/p_v\le\lambda nD\). For uniform
sampling, \(\sum_v d(v)/p_v=nD\). By Markov's inequality, for every \(0<\varepsilon<1\),
a reweighted sample of \(k\ge576^2\lambda n/(\bar d\varepsilon^2)\) vertices gives a
\((1\pm\varepsilon)\)-approximation to \(\mathrm{OPT}\) with probability at least
\(2/3\).

For metric weights, importance sampling gives an error bound relative to the optimal cut
value. For \(\lambda\)-approximate importance sampling, the bound below gains a factor
of \(\sqrt{\lambda}\).

\begin{corollary}[Relative error for metric weights]
\label{cor:metric-relative-error}
Assume \(A_{uv}\ge0\) and \(A_{uv}\le A_{uw}+A_{wv}\) for all \(u,v,w\in V\). Use exact
importance sampling \(p_v=d(v)/D\), and write \(\mathrm{OPT}=\MC(A)\). Then
\[
  \E\left|\MC(\widehat A)-\mathrm{OPT}\right|
  \le\frac{192\sqrt{2}}{\sqrt{k}}\,\mathrm{OPT}.
\]
\end{corollary}

\begin{proof}
Theorem~\ref{thm:main} bounds the expected error by \(48n^2H/\sqrt{k}\), so we need to
bound \(n^2H\) using \(\mathrm{OPT}\). Apply Cauchy--Schwarz to the sum over \(v\),
using \(\sum_vp_v=1\), then substitute \(p_v=d(v)/D\), which gives
\[
  n^2H
  =n^2\sum_v\sqrt{p_v}\left(p_v\sum_u p_uK_{uv}^2\right)^{1/2}
  \le n^2\left(\sum_{u,v}p_up_vK_{uv}^2\right)^{1/2}
  =D\left(\sum_{u,v}\frac{A_{uv}^2}{d(u)d(v)}\right)^{1/2}
  \le\sqrt{2}D.
\]
The last inequality follows from the next calculation. For fixed \(u,v\), sum the
inequality \(A_{uv}\le A_{uw}+A_{wv}\) over the \(n\) choices \(w\in V\) to get
\[
  nA_{uv}
  \le\sum_{w\in V}A_{uw}+\sum_{w\in V}A_{wv}
  =d(u)+d(v).
\]
Multiply the preceding inequality by \(\frac1n A_{uv}/(d(u)d(v))\), and sum over \(u,v\)
to get
\[
  \sum_{u,v\in V}\frac{A_{uv}^2}{d(u)d(v)}
  \le\frac1n\left(
    \sum_{u,v\in V}\frac{A_{uv}}{d(u)}
    +\sum_{u,v\in V}\frac{A_{uv}}{d(v)}
  \right)=2.
\]
Each vertex contributes \(1\) to the first sum, since
\[
  \sum_{u,v\in V}\frac{A_{uv}}{d(u)}
  =\sum_{u\in V}\frac{\sum_{v\in V}A_{uv}}{d(u)}
  =\sum_{u\in V}1=n.
\]
The second sum also equals \(n\), by symmetry. A uniformly random cut has expected value
\(D/4\), so \(D\le4\mathrm{OPT}\). Hence \(n^2H\le4\sqrt{2}\,\mathrm{OPT}\).
Substituting this into the theorem's error bound proves the claim.
\end{proof}

\section{Solution Recovery for Max-Cut}
\label{sec:solution-recovery}

We recover a nearly optimal cut of \(A\) from an optimal cut of \(\widehat A\) and the
edges to the sampled vertices. Retain the preceding notation and write \(\Cut_A(S)\) for
the weight of a cut \(S\subseteq V\).

The same quantity \(H\) that controls the value approximation also controls the expected
approximation of the recovered cut. Recall from Lemma~\ref{lem:degree-width} that \(H\le1\) whenever every
edge weight \(A_{uv}\) lies in \([0,1]\) and the sampling distribution satisfies
\eqref{eq:probability-condition}.

\begin{theorem}[Recovery from edges to the sampled vertices]
\label{thm:recovery}
For every positive probability vector \(p\in\mathbb R^V\) and every \(k\ge1\), the
algorithm below takes an optimal cut of \(\widehat A\) and the edges to the sampled
vertices \(X_1,\ldots,X_k\), and returns a cut \(S\subseteq V\) satisfying
\begin{equation}\label{eq:expected-loss}
  \E\left[\frac{\MC(A)-\Cut_A(S)}{n^2}\right]
  \le\frac{120H}{\sqrt{k}}.
\end{equation}
The expectation is over both sampling and rounding. The algorithm uses at most \(nk\)
adjacency queries.
\end{theorem}

The query bound assumes access to the probabilities \(p_v\) and independent samples from
\(p\) without further adjacency queries.

\begin{corollary}[High-probability bound]
\label{cor:recovery-tail}
Assume \(0\le A_{uv}\le1\) for every \(u,v\in V\). Then for every positive sampling
distribution satisfying \eqref{eq:probability-condition}, including uniform and
importance sampling with \(\lambda=1\),
\begin{equation}\label{eq:tail}
  \forall\,0<\varepsilon<1,\quad
  \Prb\bigl(\MC(A)-\Cut_A(S)\ge\varepsilon n^2\bigr)
  \le\frac{120}{\varepsilon\sqrt{k}}.
\end{equation}
The probability is over both sampling and rounding. Consequently, to get error
probability at most \(\delta\in(0,1)\), it suffices to take any
\(k\ge14400/(\delta^2\varepsilon^2)\).
\end{corollary}

\begin{proof}
Lemma~\ref{lem:degree-width} gives \(H\le1\) under the stated assumptions. Since
\(\MC(A)-\Cut_A(S)\ge0\), Markov's inequality applied to \eqref{eq:expected-loss} gives
\eqref{eq:tail}. Solving for \(k\) gives the sample size bound for a given \(\delta\).
\end{proof}

Under the same bounded-weight assumption, importance sampling with parameter
\(\lambda\ge1\) gives \(H\le\sqrt{\lambda}\) by Lemma~\ref{lem:degree-width}, so an
extra factor of \(\lambda\) in the sample size suffices.

\paragraph{Overview.}
We first show how an optimal cut \(a\) of the sampled graph, together with the edges to
the sampled vertices, gives a lower bound on the value of any fractional cut \(b\) of
the original graph. The matching lemma guarantees a candidate \(b\) for which this
lower bound is close to the value of \(a\). We find such a candidate by solving a linear
program. Finally, we round the resulting fractional cut, preserving its value in
expectation, and apply the preceding sampling bounds to obtain the recovery guarantee.

\subsection{Recover a fractional cut by a linear program}
\label{sec:recovery-lp}

Write \(\Delta=\Delta_K(p,P_X)\).

\begin{lemma}[Comparison using edges to the sampled vertices]
\label{lem:recovery-loss}
Let \(a\in Z(P_X)\) be an optimal cut of the sampled graph. Then for every
\(b\in Z(p)\),
\begin{equation}\label{eq:loss-controlled}
  b^\top K(p-b)
  \ge F_K(P_X)-4\Delta
    -\frac1k\sum_{i=1}^k\bigl|(Ka)_{X_i}-(Kb)_{X_i}\bigr|.
\end{equation}
\end{lemma}

\begin{proof}
Fix \(b\in Z(p)\) and choose a matching fractional cut \(b_X\in Z(P_X)\)
using Lemma~\ref{lem:profile-matching}. We first bound the two error terms
that arise when replacing \(b\) by \(b_X\) and \(p\) by \(P_X\).
Since \(a,b\in\mathcal C(p,P_X)\) and \(K=K^\top\), the lemma gives
\begin{align*}
  |(a-b)^\top K(b-b_X)|
  &\le |(b-b_X)^\top Ka|+|(b-b_X)^\top Kb|\le2\Delta,\\
  |(P_X-p)^\top Kb|&\le2\Delta.
\end{align*}
Since \(a\) attains \(F_K(P_X)\), expanding the difference of the two
cut values and substituting \(b_X\) gives
\begin{align*}
  F_K(P_X)-b^\top K(p-b)
  &=(a-b)^\top K(P_X-a-b)+(P_X-p)^\top Kb\\
  &=(a-b)^\top K(P_X-a-b_X)-(a-b)^\top K(b-b_X)
    +(P_X-p)^\top Kb\\
  &\le(a-b)^\top K(P_X-a-b_X)+4\Delta.
\end{align*}
It remains to bound the first term using edges to the sampled vertices.
Since \(a,b_X\in Z(P_X)\), we have
\(-P_X\le P_X-a-b_X\le P_X\) coordinatewise. Thus, by symmetry of \(K\),
\[
  (a-b)^\top K(P_X-a-b_X)
  \le\sum_v P_X(v)\bigl|(Ka)_v-(Kb)_v\bigr|
  =\frac1k\sum_{i=1}^k\bigl|(Ka)_{X_i}-(Kb)_{X_i}\bigr|.
\]
Substituting this bound proves the claim.
\end{proof}

Fix an optimal cut \(a\) of the sampled graph. We now minimize the error term in
\eqref{eq:loss-controlled} over \(b\in Z(p)\). The matching lemma ensures that this
minimum is small, giving the following recovery guarantee.

\begin{lemma}[Recovery by minimizing the error]
\label{lem:recovery-minimizer}
Every \(b\in Z(p)\) minimizing
\begin{equation}\label{eq:average-error}
  \frac1k\sum_{i=1}^k\bigl|(Ka)_{X_i}-(Kb)_{X_i}\bigr|
\end{equation}
satisfies
\begin{equation}\label{eq:fractional-guarantee}
  b^\top K(p-b)\ge F_K(p)-10\Delta.
\end{equation}
\end{lemma}

\begin{proof}
Lemma~\ref{lem:profile-matching}, with \(Z(P_X)\subseteq\mathcal C(p,P_X)\) and
\(K=K^\top\), gives \(b\in Z(p)\) such that
\[
  \bigl|w^\top K(a-b)\bigr|\le\Delta
  \qquad\text{for every }w\in Z(P_X).
\]

Thus,
\[
  \frac1k\sum_{i=1}^k\bigl|(Ka)_{X_i}-(Kb)_{X_i}\bigr|
  \le2\max_{w\in Z(P_X)}\bigl|w^\top K(a-b)\bigr|
  \le2\Delta.
\]

For any minimizer \(b\), Lemmas~\ref{lem:recovery-loss} and \ref{lem:value-stability}
give
\[
\begin{aligned}
  b^\top K(p-b)
  &\ge F_K(P_X)-4\Delta
    -\frac1k\sum_{i=1}^k\bigl|(Ka)_{X_i}-(Kb)_{X_i}\bigr|\\
  &\ge F_K(P_X)-6\Delta
   \ge F_K(p)-10\Delta.
\end{aligned}
\]
\end{proof}

We compute a minimizer by solving
\begin{equation}\label{eq:LP}
\begin{aligned}
  \min_{b,t}\quad &\frac1k\sum_{i=1}^k t_i\\
  \text{subject to}\quad
  &0\le b\le p,\quad t\ge0,\\
  &-t_i\le(Ka)_{X_i}-(Kb)_{X_i}\le t_i
    &&(i=1,\ldots,k).
\end{aligned}
\end{equation}

At any optimal solution \((b,t)\) of \eqref{eq:LP},
\(t_i=\bigl|(Ka)_{X_i}-(Kb)_{X_i}\bigr|\) for every \(i\), since a larger \(t_i\) could be
decreased to improve the objective. Therefore, \(b\) minimizes \eqref{eq:average-error}
and satisfies \eqref{eq:fractional-guarantee} by Lemma~\ref{lem:recovery-minimizer}.

\subsection{Round the fractional cut and count queries}
\label{sec:recovery-complexity}

The LP assigns \(b_v\) of vertex \(v\)'s weight \(p_v\) to one side. We round by
including \(v\) in \(S\) independently with probability \(b_v/p_v\in[0,1]\), the
fraction of its weight on that side.

\begin{proof}[Proof of Theorem~\ref{thm:recovery}]
Condition on the sample \(X\) and any optimal LP solution \(b\). Independence,
\(A_{vv}=0\), \(A_{uv}=n^2p_up_vK_{uv}\), and \eqref{eq:fractional-guarantee} give
\begin{equation}\label{eq:rounding-expectation}
\begin{aligned}
  \E[\Cut_A(S)\mid X,b]
  &=\sum_{u\ne v}A_{uv}\Prb(u\in S,\ v\notin S\mid X,b)\\
  &=\sum_{u,v\in V}A_{uv}
    \frac{b_u}{p_u}\left(1-\frac{b_v}{p_v}\right)\\
  &=n^2b^\top K(p-b)\ge n^2F_K(p)-10n^2\Delta.
\end{aligned}
\end{equation}

This bound holds for every optimal \(b\). Averaging over \(X\) and applying
\eqref{eq:section4-goal} gives
\[
  \E\left[\frac{\MC(A)-\Cut_A(S)}{n^2}\right]
  \le10\E_X\Delta
   \le\frac{120H}{\sqrt{k}}.
\]

The edges to the sampled vertices determine \(\widehat A\) and all LP coefficients and
can be obtained with at most \(nk\) adjacency queries. Discarded isolated vertices can
be assigned to either side.
\end{proof}

\section{Lower Bound}
\label{sec:importance-lower-bound}

We prove a matching \(\Omega(\varepsilon^{-2})\) lower bound
for importance sampling with the estimator \(\MC(\widehat A)\) defined by
\eqref{eq:estimator}. The main example
is a balanced complete tripartite graph. An imbalance of order \(\sqrt{k}\) between its
sampled parts produces an error of order \(n^2/\sqrt{k}\), and we show that this happens
with constant probability.

For every positive integer \(n\) divisible by six, define
\[
  B_n=K_{n/2,n/2},\qquad T_n=K_{n/3,n/3,n/3},\qquad
  G_{n,k}=\begin{cases}
    B_n,&1\le k<512,\\
    T_n,&k\ge512.
  \end{cases}
\]
Both graphs are simple, unweighted, and regular, so importance sampling uses
\(p_v=d(v)/D=1/n\). By \eqref{eq:estimator}, every edge in the sampled graph has weight
\(n^2/k^2\). We use the bipartite graph to cover the finitely many sample sizes
\(k<512\).

\begin{theorem}[Lower bound for importance sampling]\label{thm:importance-lower-bound}
Set $c=1/5184$ and $\varepsilon_0=1/(4\cdot512^2)$. For every
$0<\varepsilon\le\varepsilon_0$, every integer $1\le k\le c\varepsilon^{-2}$, and every
positive integer $n$ divisible by six, importance sampling on $G_{n,k}$ satisfies
\[
  \Prb\!\left(
    |\MC(\widehat A)-\MC(A)|>\varepsilon n^2
  \right)>\frac38.
\]
Thus, for this estimator under importance sampling, a worst-case guarantee of additive error $\varepsilon n^2$
with probability at least $2/3$ requires $k=\Omega(\varepsilon^{-2})$.
\end{theorem}

\begin{proof}
We first express the error on the tripartite graph in terms of its sample counts.
Lemma~\ref{lem:importance-lower-counts} then gives the required probability bound. We
handle \(k<512\) separately using the bipartite graph.

\paragraph{Compute the estimator on the tripartite graph.}
Suppose first that \(k\ge512\), so \(G_{n,k}=T_n\). Let \(N_j\) be the number of samples
from part \(j\), and let \(M\) be the largest of these counts. Thus
\[
  (N_1,N_2,N_3)\sim\operatorname{Multinomial}(k;1/3,1/3,1/3),
  \qquad M:=\max_jN_j.
\]
The sampled graph is complete tripartite with part sizes \(N_1,N_2,N_3\), even if a
vertex is sampled more than once.

An optimal cut may be chosen to keep each part on one side. Indeed, vertices within a
part have no edges between them and have identical neighborhoods. Holding the other
parts fixed, we can therefore move all vertices of this part to whichever side gives the
larger contribution, without decreasing the cut value. Among cuts that split no part, an
optimal cut separates the largest part from the other two, leaving only the edges
between the two smallest parts uncut.

An optimal cut of the sampled graph therefore has \(M(k-M)\) crossing edges. In the
original graph, each part has size \(n/3\), so an optimal cut has
\((n/3)(2n/3)=2n^2/9\) crossing edges.
Including the sampled edge weights gives
\begin{equation}\label{eq:importance-lower-estimator}
  \MC(A)=\frac{2n^2}{9},\qquad
  \MC(\widehat A)=\frac{n^2}{k^2}M(k-M).
\end{equation}
To compare these values, let \(L:=M-k/3\ge0\) denote the amount by which the largest
sample count exceeds \(k/3\), the expected count in each part. Substituting \(M=k/3+L\),
we obtain
\begin{equation}\label{eq:importance-lower-linear}
  \frac{\MC(\widehat A)-\MC(A)}{n^2}
  =\frac{M(k-M)}{k^2}-\frac29
  =\frac{L}{3k}-\frac{L^2}{k^2}.
\end{equation}

\paragraph{Identify an event on which the estimator fails.}
When \(L\le k/6\), the quadratic term in \eqref{eq:importance-lower-linear} is at most
half the linear term. Thus an imbalance of order \(\sqrt{k}\) gives an error of order
\(n^2/\sqrt{k}\). More precisely, on the event
\[
  \frac{\sqrt{k}}{12}<L\le\frac{k}{6},
\]
equation \eqref{eq:importance-lower-linear} gives
\begin{equation}\label{eq:importance-lower-error}
\begin{aligned}
  \MC(\widehat A)-\MC(A)
  &=\frac{n^2L}{k}\left(\frac13-\frac{L}{k}\right)\\
  &\ge\frac{n^2L}{6k}
  >\frac{n^2}{72\sqrt{k}}
  \ge\varepsilon n^2.
\end{aligned}
\end{equation}
The last inequality uses $k\le1/(5184\varepsilon^2)$.

Lemma~\ref{lem:importance-lower-counts} below shows that this event has probability
greater than $3/8$, completing the case $k\ge512$.

\paragraph{Cover the smaller sample sizes.}
For \(1\le k<512\), our graph is \(B_n\). We show that the estimator fails whenever the
two parts contribute different numbers of samples.

Let \(B\sim\operatorname{Bin}(k,1/2)\) count the samples from the first part. The
sampled graph is complete bipartite with part sizes \(B\) and \(k-B\). Since all edges
of a bipartite graph can be cut,
\[
  \MC(A)=\frac{n^2}{4},\qquad
  \MC(\widehat A)=\frac{n^2}{k^2}B(k-B).
\]
The error is therefore determined by the difference between the two sample counts:
\[
  \frac{|\MC(\widehat A)-\MC(A)|}{n^2}
  =\frac14-\frac{B(k-B)}{k^2}
  =\frac{(2B-k)^2}{4k^2}.
\]
If \(B\ne k/2\), then \(2B-k\) is a nonzero integer, so
\[
  \frac{|\MC(\widehat A)-\MC(A)|}{n^2}
  \ge\frac{1}{4k^2}
  >\frac{1}{4\cdot512^2}
  =\varepsilon_0\ge\varepsilon.
\]

The final sample belongs to either part with equal probability, and at most one of these
two choices makes the counts equal. Hence \(\Prb(B=k/2)\le1/2\), and the failure
probability is at least \(1/2>3/8\).
\end{proof}

The following lemma supplies the probability bound used for the tripartite graph. It
shows that, with constant probability, the largest sample count exceeds \(k/3\) by at
least a constant times \(\sqrt{k}\) but is still at most \(k/2\).

\begin{lemma}[Imbalance among three sample counts]\label{lem:importance-lower-counts}
Let \((N_1,N_2,N_3)\) have distribution \(\operatorname{Multinomial}(k;1/3,1/3,1/3)\),
and set \(M:=\max_jN_j\) and \(L:=M-k/3\). For every \(k\ge512\),
\begin{equation}\label{eq:importance-lower-imbalance}
  \Prb\!\left(\frac{\sqrt{k}}{12}<L\le\frac{k}{6}\right)
  >\frac38.
\end{equation}
\end{lemma}

\begin{proof}
We first bound the probability that \(L>\sqrt{k}/12\) from below, then subtract the
probability that \(L>k/6\). For both bounds, it is convenient to use the sum of squared
deviations of the sample counts from their mean:
\[
  R:=\sum_{j=1}^3(N_j-k/3)^2.
\]
The advantage of \(R\) is that we can compute its first two moments by counting pairs of
samples that lie in the same part.

\paragraph[Bound R in terms of L.]{Bound \(R\) in terms of \(L\).}
For this comparison, label the parts so that \(N_1=M=k/3+L\). Its term in \(R\) is
\(L^2\), so \(R\ge L^2\). For the upper bound, using \(N_2+N_3=k-M\) and expanding gives
\[
  R=6L^2-2(M-N_2)(M-N_3)\le6L^2,
\]
since both factors in the product are nonnegative.

\paragraph{Compute the first two moments.}
For \(1\le a<b\le k\), let \(I_{ab}\) indicate whether samples \(a\) and \(b\) lie in
the same part. Part \(j\) contains \(\binom{N_j}{2}\) such pairs, so
\(\sum_j\binom{N_j}{2}=\sum_{a<b}I_{ab}\). Using \(N_j^2=N_j+2\binom{N_j}{2}\) and
\(\sum_jN_j=k\), we can write
\[
  R=\sum_{j=1}^3N_j^2-\frac{k^2}{3}
   =k+2\sum_{a<b}I_{ab}-\frac{k^2}{3}.
\]
Each indicator has mean \(1/3\) and variance \(2/9\). Distinct indicators are pairwise
independent. For disjoint pairs, this follows from independence of the samples. For
pairs sharing one sample, both indicators equal \(1\) exactly when all three samples lie
in the same part, so
\[
  \Prb(I_{ab}=I_{ac}=1)=3\left(\frac13\right)^3
  =\frac19=\Prb(I_{ab}=1)\Prb(I_{ac}=1).
\]
We can therefore compute both the mean and variance of \(R\):
\[
\begin{aligned}
  \E R
  &=k+2\binom{k}{2}\frac13-\frac{k^2}{3}
   =\frac{2k}{3},\\
  \operatorname{Var}(R)
  &=4\binom{k}{2}\frac29
   =\frac{4k(k-1)}9.
\end{aligned}
\]
In particular,
\begin{equation}\label{eq:importance-lower-moments}
\begin{aligned}
  \E R^2
  &=(\E R)^2+\operatorname{Var}(R)\\
  &=\frac{4k^2}{9}+\frac{4k(k-1)}9
   \le\frac{8k^2}{9}.
\end{aligned}
\end{equation}

\paragraph{Bound the probability of the required imbalance.}
Since \(R\le6L^2\), the event \(R>k/24\) implies \(L>\sqrt{k}/12\). To bound its
probability from below, we split the expectation of \(R\) at \(k/24\). The outcomes with
\(R\le k/24\) contribute at most \(k/24\), and Cauchy--Schwarz bounds the remaining
contribution:
\[
\begin{aligned}
  \E R
  &=\E[R\mathbf{1}_{\{R\le k/24\}}]
    +\E[R\mathbf{1}_{\{R>k/24\}}]\\
  &\le\frac{k}{24}+\sqrt{\E R^2\,\Prb(R>k/24)}.
\end{aligned}
\]
Isolating \(\Prb(R>k/24)\) and substituting the moment bounds gives
\[
\begin{aligned}
  \Prb(L>\sqrt{k}/12)
  &\ge\Prb(R>k/24)
   \ge\frac{(\E R-k/24)^2}{\E R^2}\\
  &\ge\frac{(2k/3-k/24)^2}{8k^2/9}
  =\frac{225}{512}.
\end{aligned}
\]

We now bound the probability that the imbalance is too large. If \(L>k/6\), then
\(R\ge L^2>k^2/36\), so Markov's inequality gives
\[
  \Prb(L>k/6)
  \le\Prb(R>k^2/36)
  \le\frac{\E R}{k^2/36}
  =\frac{24}{k}.
\]
Subtracting this from the lower bound, and using \(k\ge512\), gives
\[
\begin{aligned}
  \Prb\!\left(\frac{\sqrt{k}}{12}<L\le\frac{k}{6}\right)
  &=\Prb(L>\sqrt{k}/12)-\Prb(L>k/6)\\
  &\ge\frac{225}{512}-\frac{24}{k}
   \ge\frac{201}{512}>\frac38.
\end{aligned}
\]
\end{proof}

\section[Sample Complexity for Max-r-Cut]{Sample Complexity for Max-$r$-Cut}
\label{sec:main}

In this section, we extend the sampling bound of
\Cref{sec:max-cut-sample-complexity} to Max-\(r\)-Cut, showing that sampling
\(O(\varepsilon^{-2}\log(2r))\) vertices achieves an additive
\(\varepsilon n^2\)-approximation to the Max-\(r\)-Cut value.
We start by setting up the necessary notation
and state our main result in \Cref{thm:rway-main}.
We then give an overview of the proof (in \Cref{sec:rway-overview}),
which is followed by the proof itself:
We express the two cut values through a common functional (in \Cref{sec:functional}),
then reduce the comparison of these two cut values
to the difference between two linear maximization problems (in \Cref{sec:stability}),
and then bound the expectation of this difference (in \Cref{sec:sampling}).

Let \(A\in\R^{V\times V}\) be a symmetric matrix with zero diagonal and nonnegative
edge weights, i.e., \(A_{uv}\ge0\) for all \(u,v\in V\).
Isolated vertices contribute zero to every cut, so we may discard them and assume \(d(v)>0\) for every
\(v\in V\).
Fix an integer \(r\ge2\). An \(r\)-way cut of \(V\) is a tuple
\((S_1,\ldots,S_r)\) of pairwise disjoint subsets whose union is \(V\).
For a symmetric matrix \(B\) with rows and columns indexed by \(V\), the value of this cut is
\[
  \sum_{1\le j<\ell\le r}\sum_{u\in S_j}\sum_{v\in S_\ell}B_{uv}.
\]
The \(r\)-way Max-Cut value \(\MC_r(B)\) is the maximum of this value
over all \(r\)-way cuts of \(V\).

Draw \(X_1,\ldots,X_k\in V\) independently with replacement according to
\(p\), where \(k\ge1\) is an integer and \(p\in\R^V\) is a positive probability vector.
Importance sampling allows \(p_v\ge\frac{d(v)}{\lambda D}\) for all \(v\in V\),
where \(\lambda\ge1\) is a parameter, and uniform sampling uses \(p_v=1/n\).

If the same vertex is sampled more than once, its copies remain distinct vertices of the
output.
The sampled graph has vertex set \([k]\), representing \(X_1,\ldots,X_k\).
Define its edge weights \(\widehat A\) and a matrix \(K\in\R^{V\times V}\) by
\begin{equation}\label{eq:rway-estimator}
  \widehat A_{ij}:=\frac{A_{X_iX_j}}{k^2p_{X_i}p_{X_j}},
  \qquad
  K_{uv}:=\frac{A_{uv}}{n^2p_up_v}.
\end{equation}
Let \(P_X\in\R^V\) be the empirical distribution of the sample,
that is \(P_X(v):=|\{i:X_i=v\}|/k\) for every \(v\in V\).
This matrix \(K\) makes it convenient to express and compare the \(r\)-way Max-Cut values
in the original and sampled graphs, after scaling by a factor of \(1/n^2\). The expressions
below differ only in whether they use the sampling distribution \(p\) or its empirical
distribution \(P_X\).
\[
\begin{aligned}
  \frac{\MC_r(A)}{n^2}
  &=\max_{V=S_1\mathbin{\dot\cup}\cdots\mathbin{\dot\cup}S_r}
    \sum_{j<\ell}\sum_{u\in S_j}\sum_{v\in S_\ell}p_up_vK_{uv},\\
  \frac{\MC_r(\widehat A)}{n^2}
  &=\max_{V=S_1\mathbin{\dot\cup}\cdots\mathbin{\dot\cup}S_r}
    \sum_{j<\ell}\sum_{u\in S_j}\sum_{v\in S_\ell}P_X(u)P_X(v)K_{uv}.
\end{aligned}
\]
The second identity uses the fact that in the sampled graph, at least one optimal cut
keeps all copies of the same vertex together. Lemma~\ref{lem:two-values} proves this
fact and both identities.

We will use the following quantity to control the sampling error through second moment
estimates.
\[
  H:=\sum_{v\in V}p_v
    \left(\sum_{u\in V}p_uK_{uv}^2\right)^{1/2}.
\]
A simple calculation (see Lemma~\ref{lem:degree-width}) shows that \(H\le1\) whenever
every edge weight \(A_{uv}\) lies in \([0,1]\) and the sampling distribution satisfies
\begin{equation}\label{eq:rway-probability-condition}
  p_v\ge\frac{d(v)}{n^2},
  \qquad\forall\,v\in V.
\end{equation}

\begin{theorem}[Sampling approximates the \(r\)-way Max-Cut value]
\label{thm:rway-main}
For every positive probability vector \(p\in\R^V\),
\begin{equation}\label{eq:main-bound}
  \E\left|
    \frac{\MC_r(\widehat A)-\MC_r(A)}{n^2}
  \right|
  \le48\sqrt2\,H\sqrt{\frac{\log(2r)}k}.
\end{equation}
\end{theorem}

Applying Markov's inequality gives the following high-probability guarantee.

\begin{corollary}[High-probability bound]
\label{cor:sample-size}
Assume \(0\le A_{uv}\le1\) for every \(u,v\in V\). Then for every positive sampling
distribution satisfying \eqref{eq:rway-probability-condition}, including uniform and
importance sampling with \(\lambda=1\),
\begin{equation}\label{eq:rway-tail-main}
  \forall\,0<\varepsilon<1,\quad
  \Prb\left(
    |\MC_r(\widehat A)-\MC_r(A)|\ge\varepsilon n^2
  \right)
  \le\frac{48\sqrt2}{\varepsilon}\sqrt{\frac{\log(2r)}k}.
\end{equation}
Consequently, to get error probability at most \(\delta\in(0,1)\), it suffices to take
any
\begin{equation}\label{eq:sample-size}
  k\ge\frac{4608\log(2r)}{\delta^2\varepsilon^2}.
\end{equation}
The probability bound holds for each fixed \(r\ge2\).
\end{corollary}

\begin{proof}
\Cref{sec:sampling-width} verifies the bound \(H\le1\) for sampling that satisfies
\eqref{eq:rway-probability-condition}, and then Markov's inequality implies
(\ref*{eq:rway-tail-main}). Solving for \(k\) gives the sample size bound for given
\(\delta\).
\end{proof}

A simple calculation shows that, for importance sampling with parameter \(\lambda\ge1\),
an extra factor of \(\lambda\) in the sample size suffices, e.g.,
\(k\ge41472\lambda\varepsilon^{-2}\log(2r)\) gives error probability at most \(1/3\).

\subsection[Proof overview]{Proof overview}
\label{sec:rway-overview}

The proof follows the same three stages as for Max-Cut: express both optima through a
common functional, compare fractional cuts by minimax, and bound the resulting sampling
error. The main change is that each vertex's weight is now divided among \(r\) parts.
We use vector contraction to handle these assignments and Massart's finite-class lemma
to bound the maximum over the parts, introducing a factor of \(\sqrt{\log(2r)}\) in the
error.

Our first step is to view sampling as a change in vertex weights. Given \(K\) and a
probability vector \(q\in\R^V\), we define \(F_{K,r}(q)\) as the maximum \(r\)-way cut
value with respect to \(K\) and vertex weights \(q_v\), meaning that each edge \((u,v)\)
has weight \(q_uq_vK_{uv}\). It will be convenient to allow fractional cuts, where each
vertex's weight \(q_v\) may be split among the \(r\) parts, although we show in
\cref{sec:functional} that this does not change the optimum value, and in fact
\[
  F_{K,r}(p)=\frac{\MC_r(A)}{n^2},
  \qquad F_{K,r}(P_X)=\frac{\MC_r(\widehat A)}{n^2}.
\]

We then compare in \cref{sec:stability} these two values, and more generally
\(F_{K,r}(q)\) and \(F_{K,r}(s)\) for any two probability vectors \(q,s\in\R^V\). We do
this by matching fractional cuts of the two vertex weights \(q\) and \(s\). For either
\(t\in\{q,s\}\), a fractional cut is represented by \(a=(a^1,\ldots,a^r)\in(\R^V)^r\)
with \(a^j\ge0\) and \(\sum_j a^j=t\), and has value
\(\frac12\sum_{j\ne\ell}(a^j)^\top Ka^\ell\). We relax the doubled cut value
\(\sum_{j\ne\ell}(a^j)^\top Ka^\ell\) to \(\sum_{j\ne\ell}(a^j)^\top Kc^\ell\),
allowing \(c\in(\R^V)^r\) to vary independently of \(a\) with \(c^j\ge0\) and either
\(\sum_j c^j=q\) or \(\sum_j c^j=s\). For fixed \(c\), maximizing over \(a\) gives
\(M_{t,r}(c):=\max_{\substack{a^1,\ldots,a^r\ge0\\\sum_j a^j=t}}
\sum_{j\ne\ell}(a^j)^\top Kc^\ell\), and our analysis compares \(M_{q,r}(c)\) and
\(M_{s,r}(c)\) for the same \(c\). Let \(\Delta_{K,r}(q,s)\) be their largest absolute
difference over a suitable region \(\mathcal C_r(q,s)\) of vectors \(c\), i.e.,
\[
  \Delta_{K,r}(q,s):=
  \max_{c\in\mathcal C_r(q,s)}|M_{q,r}(c)-M_{s,r}(c)|.
\]

A minimax argument converts the comparison of these maxima into a matching of fractional
cuts. For each \(a\in(\R^V)^r\) with \(a^j\ge0\) and \(\sum_j a^j=q\), it produces
a matching \(b\in(\R^V)^r\), \(b^j\ge0\), \(\sum_j b^j=s\), such that
\(\sum_{j\ne\ell}(a^j)^\top Kc^\ell\) and \(\sum_{j\ne\ell}(b^j)^\top Kc^\ell\)
are close for all feasible \(c\) at once, meaning that
\[
  \max_{c\in\mathcal C_r(q,s)}
  \left|\sum_{j\ne\ell}(a^j-b^j)^\top Kc^\ell\right|
  \le\Delta_{K,r}(q,s).
\]

Using this matching, a simple calculation shows that
\[
  |F_{K,r}(q)-F_{K,r}(s)|\le\Delta_{K,r}(q,s).
\]

Applying this bound with \(q=p\) and \(s=P_X\) reduces the problem to bounding
\(\E_X\Delta_{K,r}(p,P_X)\), which we address in \cref{sec:sampling}. The main
difficulty is that both the sampling error for a given \(c\) and the feasible region
for \(c\) depend on the sample \(X\). The maximizing vector can therefore adapt to the
sample, so a bound for each \(c\) fixed in advance does not directly control
\(\Delta_{K,r}(p,P_X)\). Similar difficulties arise in statistical learning when
comparing a hypothesis's error on the sample used to choose it with its error under
the underlying distribution. We borrow from that domain a standard tool called
\emph{symmetrization}. We draw an independent \emph{ghost sample} \(Y_1,\ldots,Y_k\)
from \(p\), solely for the sake of analysis, and denote its empirical distribution by
\(P_Y\) and use it to rewrite \(\Delta_{K,r}(p,P_X)\) as
\[
  \Delta_{K,r}(p,P_X)
  =\max_{c\in\mathcal C_r(p,P_X)}
    \left|\E_Y\bigl[M_{P_X,r}(c)-M_{P_Y,r}(c)\bigr]\right|.
\]

We enlarge the feasible region for \(c\) to one that is invariant under swaps between
\(X_i\) and \(Y_i\), and then draw \(k\) independent random signs
\(\tau_1,\ldots,\tau_k\in\{\pm1\}\), where \(+1\) leaves \(X_i,Y_i\) in place, and
\(-1\) exchanges them. Crucially, the enlarged region does not change under these
exchanges. Writing \(g_c(v):=\max_{1\le j\le r}\sum_{\ell\ne j}(Kc^\ell)_v\), the
swaps turn \(M_{P_X,r}(c)-M_{P_Y,r}(c)\) into
\[
  \frac1k\sum_{i=1}^k\tau_i\bigl(g_c(X_i)-g_c(Y_i)\bigr).
\]
This lets us apply Maurer's vector contraction inequality to reduce to a signed sum
linear in \(c\). We can therefore remove the maximum over \(c\) by bounding each
\(\sum_j|c_v^j|\) by its worst-case value.
A simple second moment calculation then gives the required bound, with
Massart's finite-class lemma contributing a factor of \(\sqrt{\log(2r)}\)
to control the maximum over the \(r\) parts:
\[
  \E_X\Delta_{K,r}(p,P_X)\le48\sqrt2\,H\sqrt{\frac{\log(2r)}k}.
\]

\subsection[Reduce both r-way Max-Cut values to one functional]{Reduce both \(r\)-way Max-Cut values to one functional}
\label{sec:functional}

We express the normalized \(r\)-way Max-Cut values of the original graph \(A\) and the
sampled graph \(\widehat A\) as evaluations of the same functional \(F_{K,r}(\cdot)\) at
\(p\) and \(P_X\), respectively.

To define \(F_{K,r}\), we first describe fractional cuts. Cuts are usually represented by
\(x_v^j\in\{0,1\}\), where vertices \(v\) with \(x_v^j=1\) are in part \(j\), and
\(\sum_{j=1}^r x_v^j=1\) for every \(v\). Allowing \(x_v^j\in[0,1]\) gives a fractional
cut, in which a fraction \(x_v^j\) of vertex \(v\) goes to part \(j\), and these fractions
still sum to one. With vertex weights \(q_v\ge0\), the weight of \(v\) placed in part
\(j\) is \(a_v^j=q_vx_v^j\), which we call the \emph{weighted fraction} of \(v\) in that
part. For simplicity, we will simply refer to \(a_v^j\) as the fraction of \(v\) in that
part. Thus a fractional cut has \(a_v^j\ge0\) and \(\sum_j a_v^j=q_v\), and an ordinary cut has \(a_v^j\in\{0,q_v\}\).
The probability vectors \(p\) and \(P_X\) provide the vertex weights for the original
and sampled graphs respectively. Using fractional cuts in a vertex-weighted graph
lets us express both cut values through the same functional.

Define the feasible fractional cuts by
\begin{equation}\label{eq:partitions}
  Z_r(q):=\left\{a=(a^1,\ldots,a^r)\in(\R^V)^r:
    a_v^j\ge0,\ \sum_{j=1}^r a_v^j=q_v\text{ for every }v\in V
  \right\}
\end{equation}
and their optimal value by
\begin{equation}\label{eq:functional}
  F_{K,r}(q):=\max_{a\in Z_r(q)}
    \sum_{j<\ell}(a^j)^\top Ka^\ell
  =\frac12\max_{a\in Z_r(q)}
    \sum_{j\ne\ell}(a^j)^\top Ka^\ell.
\end{equation}
The tuple \(a=(a^1,\ldots,a^r)\) collects the weighted fractions. Here and below, sums over
vertices range over \(V\), and part indices \(j,\ell\) range from \(1\) to \(r\).
The contribution between parts \(j\) and \(\ell\) is
\(\sum_{u,v}a_u^jK_{uv}a_v^\ell=(a^j)^\top Ka^\ell\). The sum \(\sum_{j\ne\ell}\) is over
ordered pairs of distinct parts, so symmetry of \(K\) gives the second equality, with
\(1/2\) removing double counting.
Allowing fractional cuts makes the feasible set convex, which is useful for the
comparison in \cref{sec:stability}.

\begin{lemma}[The maximum is attained by a cut]
\label{lem:corners}
For every nonnegative weight vector \(q\in\R^V\), there exists a maximizer
\(a\in Z_r(q)\) in \eqref{eq:functional} that corresponds to a cut, i.e.,
\(a_v^j\in\{0,q_v\}\) for all \(v\in V\) and \(1\le j\le r\).
\end{lemma}

\begin{proof}
Since \(K_{vv}=0\), the objective \(\sum_{j<\ell}(a^j)^\top Ka^\ell\) is affine in each
vertex's assignment \((a_v^1,\ldots,a_v^r)\) (when the other assignments are fixed).
An affine function on the simplex \(\{z\in\R^r:z_j\ge0,\ \sum_{j=1}^r z_j=q_v\}\)
attains its maximum at a vertex, so we can move \((a_v^1,\ldots,a_v^r)\) to a vertex
without decreasing the objective. At such a vertex, one coordinate equals \(q_v\)
and the others equal zero. Starting from a maximizer and applying this operation
iteratively yields a maximizer that has \(a_v^j\in\{0,q_v\}\) for every \(v,j\).
\end{proof}

\begin{lemma}[The two \(r\)-way Max-Cut values]
\label{lem:two-values}
With \(p\), \(P_X\), and \(K\) defined above,
\[
  F_{K,r}(p)=\frac{\MC_r(A)}{n^2},
  \qquad F_{K,r}(P_X)=\frac{\MC_r(\widehat A)}{n^2}.
\]
\end{lemma}

\begin{proof}
For the first identity, Lemma~\ref{lem:corners} lets us maximize over \(r\)-way cuts
\(V=S_1\mathbin{\dot\cup}\cdots\mathbin{\dot\cup}S_r\). Since
\(p_up_vK_{uv}=A_{uv}/n^2\),
\[
  F_{K,r}(p)
  =\max_{V=S_1\mathbin{\dot\cup}\cdots\mathbin{\dot\cup}S_r}
    \sum_{j<\ell}\sum_{u\in S_j}\sum_{v\in S_\ell}p_up_vK_{uv}
  =\frac{\MC_r(A)}{n^2}.
\]

For the second identity, we first show that all copies of a repeatedly sampled vertex
can be moved to the same part without decreasing the cut value. Start with an optimal
cut for the sampled subgraph (induced on \(X_1,\ldots,X_k\)).

For an original vertex \(v\), its sampled copies have no edges between them and have the
same edge weights to all other sampled vertices. Holding those other vertices fixed,
choose a part maximizing the cut contribution of one copy of \(v\). The same choice
maximizes the contribution of every copy. If several parts tie, choose any of them
arbitrarily. Put all copies in that part and repeat for each original vertex. These
moves cannot decrease the cut value, so the final cut is still optimal and keeps all
copies of each vertex together.

Such a cut is described by a partition \(V=S_1\mathbin{\dot\cup}\cdots\mathbin{\dot\cup}S_r\).
All copies of vertices in \(S_j\) lie in part \(j\). For \(u\in S_j\) and \(v\in S_\ell\) with \(j<\ell\),
there are
\(kP_X(u)\,kP_X(v)\) crossing pairs of copies. Each pair \((i,i')\) with \(X_i=u\) and
\(X_{i'}=v\) contributes to the cut value divided by \(n^2\) as follows:
\[
  \frac{\widehat A_{ii'}}{n^2}
  =\frac{A_{uv}}{k^2n^2p_up_v}
  =\frac{K_{uv}}{k^2}.
\]
Therefore, applying Lemma~\ref{lem:corners} with \(q=P_X\),
\begin{equation}\label{eq:sample-objective}
  F_{K,r}(P_X)
  =\max_{V=S_1\mathbin{\dot\cup}\cdots\mathbin{\dot\cup}S_r}
    \sum_{j<\ell}\sum_{u\in S_j}\sum_{v\in S_\ell}P_X(u)P_X(v)K_{uv}
  =\frac{\MC_r(\widehat A)}{n^2}.
\end{equation}
\end{proof}

By Lemma~\ref{lem:two-values},
\[
  \frac{|\MC_r(\widehat A)-\MC_r(A)|}{n^2}
  =\left|F_{K,r}(P_X)-F_{K,r}(p)\right|.
\]

\subsection{Compare cut values under two vertex distributions}
\label{sec:stability}

In this subsection, we bound the difference between \(F_{K,r}(q)\) and \(F_{K,r}(s)\)
for two probability vectors \(q,s\) on \(V\). In our intended application, we will use
\(q=p\) and \(s=P_X\).

For a nonnegative vector \(q\) and a tuple \(c\in(\R^V)^r\), let \(M_{q,r}(c)\) be the
maximum of \(\sum_{j\ne\ell}(x^j)^\top Kc^\ell\) over \(x\in Z_r(q)\), namely
\begin{equation}\label{eq:linear-value}
  M_{q,r}(c):=
  \max_{x\in Z_r(q)}\sum_{j\ne\ell}(x^j)^\top Kc^\ell
  =\sum_{v\in V}q_vg_c(v),
\end{equation}
where
\begin{equation}\label{eq:vertex-score}
  g_c(v):=\max_{1\le j\le r}\sum_{\ell\ne j}(Kc^\ell)_v.
\end{equation}
The equality follows by choosing \(x_v^j=q_v\) for a part \(j\) attaining \(g_c(v)\),
and \(x_v^\ell=0\) for all \(\ell\ne j\).

To compare \(M_{q,r}\) and \(M_{s,r}\), we use a larger domain that contains both
\(Z_r(q)\) and \(Z_r(s)\) and is symmetric and convex.
\begin{equation}\label{eq:directions}
  \mathcal C_r(q,s):=\conv\bigl(
    Z_r(q)\cup-Z_r(q)\cup Z_r(s)\cup-Z_r(s)\bigr).
\end{equation}

We consider their largest absolute difference, defined by
\begin{equation}\label{eq:discrepancy}
  \Delta_{K,r}(q,s):=
  \max_{c\in\mathcal C_r(q,s)}|M_{q,r}(c)-M_{s,r}(c)|.
\end{equation}

\begin{lemma}[Comparison of cut values]
\label{lem:rway-value-stability}
For every pair of probability vectors \(q,s\in\R^V\),
\begin{equation}\label{eq:stability}
  |F_{K,r}(q)-F_{K,r}(s)|\le\Delta_{K,r}(q,s).
\end{equation}
\end{lemma}

Taking \(q=p\) and \(s=P_X\), this lemma gives
\(|F_{K,r}(P_X)-F_{K,r}(p)|\le\Delta_{K,r}(p,P_X)\).
It then remains to bound \(\E_X\Delta_{K,r}(p,P_X)\), which we do in
\cref{sec:sampling}.

To prove Lemma~\ref{lem:rway-value-stability}, we first prove a bound that matches
fractional cuts under \(q\) and \(s\).

\begin{lemma}[Matching fractional cuts]
\label{lem:match}
Let \(q,s\in\R^V\) be probability vectors. For every \(a\in Z_r(q)\), there exists
\(b\in Z_r(s)\) such that
\begin{equation}\label{eq:match}
  \max_{c\in\mathcal C_r(q,s)}
  \left|\sum_{j\ne\ell}(a^j-b^j)^\top Kc^\ell\right|
  \le\Delta_{K,r}(q,s).
\end{equation}
Conversely, every \(b\in Z_r(s)\) has a matching \(a\in Z_r(q)\) satisfying the same
bound.
\end{lemma}

\begin{proof}
Fix \(a\in Z_r(q)\). Using \(\mathcal C_r(q,s)=-\mathcal C_r(q,s)\) and applying Sion's
minimax theorem \cite{sion} to the function
\((b,c)\mapsto\sum_{j\ne\ell}(a^j-b^j)^\top Kc^\ell\), which is affine in each variable,
on the compact convex sets \(Z_r(s)\) and \(\mathcal C_r(q,s)\), we obtain
\begin{align*}
  &\min_{b\in Z_r(s)}\max_{c\in\mathcal C_r(q,s)}
    \left|\sum_{j\ne\ell}(a^j-b^j)^\top Kc^\ell\right|\\
  &\quad=\min_{b\in Z_r(s)}\max_{c\in\mathcal C_r(q,s)}
    \sum_{j\ne\ell}(a^j-b^j)^\top Kc^\ell\\
  &\quad=\max_{c\in\mathcal C_r(q,s)}\min_{b\in Z_r(s)}
    \sum_{j\ne\ell}(a^j-b^j)^\top Kc^\ell\\
  &\quad=\max_{c\in\mathcal C_r(q,s)}
    \left(\sum_{j\ne\ell}(a^j)^\top Kc^\ell-M_{s,r}(c)\right)\\
  &\quad\le\max_{c\in\mathcal C_r(q,s)}
    \bigl(M_{q,r}(c)-M_{s,r}(c)\bigr)
   \le\Delta_{K,r}(q,s).
\end{align*}
The minimum is attained, so \eqref{eq:match} follows.

Interchanging \(q\) and \(s\) gives the same bound for every \(b\in Z_r(s)\), proving
the converse.
\end{proof}

\begin{proof}[Proof of Lemma~\ref{lem:rway-value-stability}]
Let \(a\in Z_r(q)\) attain \(F_{K,r}(q)\), and choose \(b\in Z_r(s)\) as in
Lemma~\ref{lem:match}. Write \(\Delta=\Delta_{K,r}(q,s)\).

For every \(c\in\mathcal C_r(q,s)\), Lemma~\ref{lem:match} gives
\[
  \left|\sum_{j\ne\ell}(a^j-b^j)^\top Kc^\ell\right|\le\Delta.
\]
Since \(F_{K,r}(s)\ge\frac12\sum_{j\ne\ell}(b^j)^\top Kb^\ell\), the symmetry of \(K\) gives
\begin{equation}\label{eq:expanded-difference}
\begin{aligned}
  F_{K,r}(q)-F_{K,r}(s)
  &\le\frac12\sum_{j\ne\ell}
       \bigl((a^j)^\top Ka^\ell-(b^j)^\top Kb^\ell\bigr)\\
  &=\frac12\sum_{j\ne\ell}(a^j-b^j)^\top Ka^\ell
    +\frac12\sum_{j\ne\ell}(a^j-b^j)^\top Kb^\ell\\
  &\le\frac12\left|\sum_{j\ne\ell}(a^j-b^j)^\top Ka^\ell\right|
    +\frac12\left|\sum_{j\ne\ell}(a^j-b^j)^\top Kb^\ell\right|.
\end{aligned}
\end{equation}
Since \(a\in Z_r(q)\) and \(b\in Z_r(s)\), both lie in \(\mathcal C_r(q,s)\), and we can
apply \eqref{eq:match} to get
\[
  F_{K,r}(q)-F_{K,r}(s)\le\frac12\Delta+\frac12\Delta=\Delta.
\]
Interchanging \(q\) and \(s\) gives the same bound for \(F_{K,r}(s)-F_{K,r}(q)\),
proving \eqref{eq:stability}.
\end{proof}

\subsection{Bound the sampling error}
\label{sec:sampling}

In this subsection we prove that
\begin{equation}\label{eq:sampling-target}
  \E_X\Delta_{K,r}(p,P_X)\le48\sqrt2\,H\sqrt{\frac{\log(2r)}k}.
\end{equation}
Together with Lemmas~\ref{lem:two-values} and~\ref{lem:rway-value-stability}, this
completes the proof of Theorem~\ref{thm:rway-main}. Recall that \(\Delta_{K,r}(p,P_X)\)
is the largest absolute difference between the exact and sample averages
\begin{equation}\label{eq:population-sample-scores}
  M_{p,r}(c)=\sum_vp_v\max_{1\le j\le r}\sum_{\ell\ne j}(Kc^\ell)_v,
  \qquad
  M_{P_X,r}(c)=\frac1k\sum_{i=1}^k\max_{1\le j\le r}\sum_{\ell\ne j}(Kc^\ell)_{X_i},
\end{equation}
over \(c\in\mathcal C_r(p,P_X)\). The challenge is that this domain depends on the same
sample \(X\) used in \(M_{P_X,r}\).

The proof has two steps, using several tools from statistical learning theory.

\begin{enumerate}
  \item \textbf{Symmetrization.}
  Instead of comparing \(P_X\) with its expectation \(p\), we compare the empirical
  distributions \(P_X,P_Y\) of two independent samples \(X,Y\). We enlarge the domain of
  \(c\) to a region \(\mathcal B^{(r)}_{X,Y}\).

  We use symmetrization and Maurer's vector contraction inequality
  \cite{maurer} to reduce the problem to bounding the expected maximum of sums of
  \((Kc^j)_{X_i}\) with random signs\nobreakspace(Lemma~\ref{lem:linear-reduction}).

  \item \textbf{Second-moment bounds.}
  The signed sums are linear in \(c\), so we bound their maximum over the
  region one vertex at a time. We bound the maximum by three sums, weighted by
  \(p\), \(P_X\), and \(P_Y\). We bound each sum's expectation by
  \(2\sqrt2\,H\sqrt{\log(2r)/k}\) using the upper \(L^1\) Khintchine bound,
  Massart's finite-class lemma, and Jensen's inequality, obtaining a total of at most
  \(6\sqrt2\,H\sqrt{\log(2r)/k}\)
  (Lemma~\ref{lem:signed-maximum}).
\end{enumerate}
Combining the two estimates gives \eqref{eq:sampling-target}.
Lemma~\ref{lem:rway-value-stability} then bounds the expected normalized \(r\)-way
Max-Cut error by \(48\sqrt2\,H\sqrt{\log(2r)/k}\).

\subsection*{Step 1. Symmetrization}

\begin{lemma}[Reduction to a signed linear maximum]
\label{lem:linear-reduction}
Let \(Y_1,\ldots,Y_k\) be independent draws according to \(p\), independent of \(X\),
with sample-frequency vector \(P_Y\). Let \(\sigma_{ij}\in\{\pm1\}\), for
\(1\le i\le k\) and \(0\le j\le r\), be independent uniform signs, independent of
\(X,Y\), and define
\begin{equation}\label{eq:enlargement}
  \mathcal B^{(r)}_{X,Y}:=\left\{c\in(\R^V)^r:
    \sum_{j=1}^r|c_v^j|\le p_v+P_X(v)+P_Y(v)
    \text{ for every }v\in V\right\}.
\end{equation}
Then
\begin{equation}\label{eq:linear-reduction}
  \E_X\Delta_{K,r}(p,P_X)
  \le8\E_{X,Y,\sigma}\max_{c\in\mathcal B^{(r)}_{X,Y}}
  \left|\frac1k\sum_{i=1}^k\sum_{j=1}^r
    (\sigma_{i0}+\sigma_{ij})(Kc^j)_{X_i}\right|.
\end{equation}
\end{lemma}

\begin{proof}
We begin with the symmetrization step using a ghost sample. For every
\(c\in\mathcal C_r(p,P_X)\), the independent sample average has mean \(M_{p,r}(c)\), since
\[
  \E_YM_{P_Y,r}(c)=\sum_vp_vg_c(v)=M_{p,r}(c).
\]
The region \(\mathcal B^{(r)}_{X,Y}\) is convex and contains each of the four sets
defining \(\mathcal C_r(p,P_X)\), so
\begin{equation}\label{eq:containment}
  \mathcal C_r(p,P_X)\subseteq\mathcal B^{(r)}_{X,Y}.
\end{equation}

With \(X\) fixed, \(\mathcal C_r(p,P_X)\) does not depend on \(Y\), the absolute value
of an expectation is at most the expectation of the absolute value and the maximum of
expectations is at most the expectation of the maximum (Jensen's inequality). Thus
\begin{align*}
  \Delta_{K,r}(p,P_X)
  &=\max_{c\in\mathcal C_r(p,P_X)}
    \left|\E_Y\bigl[M_{P_X,r}(c)-M_{P_Y,r}(c)\bigr]\right|\\
  &\le\max_{c\in\mathcal C_r(p,P_X)}
    \E_Y|M_{P_X,r}(c)-M_{P_Y,r}(c)|\\
  &\le\E_Y\max_{c\in\mathcal C_r(p,P_X)}
    |M_{P_X,r}(c)-M_{P_Y,r}(c)|\\
  &\le\E_Y\max_{c\in\mathcal B^{(r)}_{X,Y}}
    \left|\frac1k\sum_{i=1}^k\bigl(g_c(X_i)-g_c(Y_i)\bigr)\right|.
\end{align*}

We next take expectation over \(X\) and use pairwise exchanges to express the expected
maximum in terms of sums with random signs (Rademacher symmetrization). Let
\(\tau_1,\ldots,\tau_k\in\{\pm1\}\) be independent uniform signs, independent of
\(X,Y,\sigma\). For each fixed sign vector \(\tau\in\{\pm1\}^k\), set
\[
  (\widetilde X_i,\widetilde Y_i):=
  \begin{cases}
    (X_i,Y_i),&\tau_i=1,\\
    (Y_i,X_i),&\tau_i=-1.
  \end{cases}
\]
Each exchange reverses its pair's contribution, so
\[
  g_c(\widetilde X_i)-g_c(\widetilde Y_i)
  =\tau_i\bigl(g_c(X_i)-g_c(Y_i)\bigr).
\]
The exchanges preserve the joint distribution of the samples. They also preserve
\(P_X+P_Y\), and thus
\(\mathcal B^{(r)}_{\widetilde X,\widetilde Y}
=\mathcal B^{(r)}_{X,Y}\).

Together with the preceding bound, these identities give
\begin{align}
  \E_X\Delta_{K,r}(p,P_X)
  &\le\E_{X,Y}\max_{c\in\mathcal B^{(r)}_{X,Y}}
    \left|\frac1k\sum_{i=1}^k\bigl(g_c(X_i)-g_c(Y_i)\bigr)\right|\notag\\
  &=\E_{X,Y,\tau}\max_{c\in\mathcal B^{(r)}_{X,Y}}
    \left|\frac1k\sum_{i=1}^k\tau_i\bigl(g_c(X_i)-g_c(Y_i)\bigr)\right|\notag\\
  &\le2\E_{X,Y,\tau}\max_{c\in\mathcal B^{(r)}_{X,Y}}
    \left|\frac1k\sum_{i=1}^k\tau_i g_c(X_i)\right|.
    \label{eq:symmetrization}
\end{align}
The last inequality uses the triangle inequality and symmetry between \(X\) and \(Y\).

Finally, \(\phi:(z_0,z_1,\ldots,z_r)\mapsto z_0-\min_{1\le j\le r}z_j\)
is \(\sqrt2\)-Lipschitz with respect to the Euclidean norm and vanishes at \(0\).
For fixed \(X,Y\), write \(g_c(v)=\phi(f_c(v))\), where
\[
  f_c(v):=\left(\sum_{\ell=1}^r(Kc^\ell)_v,
    (Kc^1)_v,\ldots,(Kc^r)_v\right).
\]
The region \(\mathcal B^{(r)}_{X,Y}\) is compact and contains zero. Since \(f_c\)
depends linearly on \(c\), the tuples \((f_c(X_1),\ldots,f_c(X_k))\) also form a compact
set containing zero. The absolute-value form of Maurer's vector contraction inequality
(Lemma~\ref{lem:absolute-vector-contraction}; see \cite[Corollary~1]{maurer})
therefore gives
\begin{equation}\label{eq:conditional-linear-reduction}
  \E_\tau\max_{c\in\mathcal B^{(r)}_{X,Y}}
    \left|\frac1k\sum_{i=1}^k\tau_i g_c(X_i)\right|
  \le4\E_\sigma\max_{c\in\mathcal B^{(r)}_{X,Y}}
    \left|\frac1k\sum_{i=1}^k\sum_{j=1}^r
      (\sigma_{i0}+\sigma_{ij})(Kc^j)_{X_i}\right|.
\end{equation}
See \cref{app:contraction} for the application to \(g_c\).

Taking expectation over \(X,Y\) and substituting into \eqref{eq:symmetrization}
proves the lemma.

\end{proof}

\subsection*{Step 2. Second-moment bounds}
\label{sec:signed-bound}

\begin{lemma}[Bound for the signed maximum]
\label{lem:signed-maximum}
For the samples and signs above,
\begin{equation}\label{eq:signed-maximum}
  \E_{X,Y,\sigma}\max_{c\in\mathcal B^{(r)}_{X,Y}}
  \left|\frac1k\sum_{i=1}^k\sum_{j=1}^r
    (\sigma_{i0}+\sigma_{ij})(Kc^j)_{X_i}\right|
  \le6\sqrt2\,H\sqrt{\frac{\log(2r)}k}.
\end{equation}
\end{lemma}

\begin{proof}
We first bound the maximum. Fix \(X,Y,\sigma\), and let \(b_v\) denote the bound on
\(\sum_j|c_v^j|\) in the definition of \(\mathcal B^{(r)}_{X,Y}\), so
\[
  b_v:=p_v+P_X(v)+P_Y(v).
\]
Since \(K\) is symmetric, every \(c\in\mathcal B^{(r)}_{X,Y}\) satisfies
\begin{equation}\label{eq:linear-maximum-by-rows}
\begin{aligned}
  \left|\sum_{i=1}^k\sum_{j=1}^r
    (\sigma_{i0}+\sigma_{ij})(Kc^j)_{X_i}\right|
  &=\left|\sum_v\sum_{j=1}^r c_v^j
    \sum_{i=1}^k(\sigma_{i0}+\sigma_{ij})K_{vX_i}\right|\\
  &\le\sum_vb_v\max_{1\le j\le r}
    \left|\sum_{i=1}^k(\sigma_{i0}+\sigma_{ij})K_{vX_i}\right|.
\end{aligned}
\end{equation}
Here we used \(\sum_j|c_v^j|\le b_v\) at each vertex.

Taking the maximum over \(c\in\mathcal B^{(r)}_{X,Y}\) and substituting the definition of
\(b_v\), we obtain
\begin{equation}\label{eq:rway-three-weight-split}
\begin{aligned}
  &\max_{c\in\mathcal B^{(r)}_{X,Y}}
    \left|\sum_{i=1}^k\sum_{j=1}^r
      (\sigma_{i0}+\sigma_{ij})(Kc^j)_{X_i}\right|
   \le\sum_vp_v\max_j
    \left|\sum_i(\sigma_{i0}+\sigma_{ij})K_{vX_i}\right|\\
  &\qquad+\sum_vP_X(v)\max_j
    \left|\sum_i(\sigma_{i0}+\sigma_{ij})K_{vX_i}\right|
    +\sum_vP_Y(v)\max_j
    \left|\sum_i(\sigma_{i0}+\sigma_{ij})K_{vX_i}\right|.
\end{aligned}
\end{equation}
We have bounded the maximum over all \(c\) by three explicit sums. We bound them
separately because the weights from \(P_X\) depend on the signed sums, whereas the
weights from \(P_Y\) are independent of them.

\paragraph[Bound the sum weighted by p.]{Bound the sum weighted by \(p\).}
Since \(K\) is symmetric, we can write \(H=\sum_vp_vh_v\), where, for each \(v\in V\),
\[
  h_v:=\left(\sum_up_uK_{vu}^2\right)^{1/2}.
\]
For fixed \(X\), the upper \(L^1\) Khintchine inequality
(Lemma~\ref{lem:upper-l1-khintchine}) gives
\[
  \E_\sigma\left|\sum_i\sigma_{i0}K_{vX_i}\right|
  \le\left(\sum_iK_{vX_i}^2\right)^{1/2}.
\]
The triangle inequality and Massart's finite-class lemma
(Lemma~\ref{lem:massart-finite-class}) therefore give
\begin{equation}\label{eq:label-maximum}
\begin{aligned}
  &\E_\sigma\max_{1\le j\le r}
    \left|\sum_i(\sigma_{i0}+\sigma_{ij})K_{vX_i}\right|
   \le
    \E_\sigma\left|\sum_i\sigma_{i0}K_{vX_i}\right|
    +\E_\sigma\max_{1\le j\le r}
      \left|\sum_i\sigma_{ij}K_{vX_i}\right|\\
  &\quad\le\left(1+\sqrt{2\log(2r)}\right)
    \left(\sum_iK_{vX_i}^2\right)^{1/2}
    \le2\sqrt{2\log(2r)}
    \left(\sum_iK_{vX_i}^2\right)^{1/2}.
\end{aligned}
\end{equation}

Taking expectation over \(X\) and applying Jensen's inequality gives
\[
\begin{aligned}
  \E_{X,\sigma}\max_j
    \left|\sum_i(\sigma_{i0}+\sigma_{ij})K_{vX_i}\right|
  &\le2\sqrt{2\log(2r)}\,
    \E_X\left(\sum_iK_{vX_i}^2\right)^{1/2}\\
  &\le2\sqrt{2\log(2r)}
    \left(\sum_i\E_XK_{vX_i}^2\right)^{1/2}
   =2\sqrt{2k\log(2r)}\,h_v,
\end{aligned}
\]
where the last equality uses \(\E_XK_{vX_i}^2=\sum_up_uK_{vu}^2=h_v^2\), since each
\(X_i\) is drawn according to \(p\).

We thus obtain that
\begin{equation}\label{eq:p-term}
  \E_{X,\sigma}\sum_vp_v\max_j
    \left|\sum_i(\sigma_{i0}+\sigma_{ij})K_{vX_i}\right|
  \le2\sqrt{2k\log(2r)}\sum_vp_vh_v
   =2H\sqrt{2k\log(2r)}.
\end{equation}

\paragraph[Bound the sum weighted by P_Y.]{Bound the sum weighted by \(P_Y\).}
We can use the preceding estimate directly for \(P_Y\), because the sample \(Y\) is
independent of \(X\) and \(\sigma\). Since \(\E_YP_Y(v)=p_v\), equation
\eqref{eq:p-term} gives
\begin{equation}\label{eq:PY-term}
\begin{aligned}
  &\E_{X,Y,\sigma}\sum_vP_Y(v)\max_j
    \left|\sum_i(\sigma_{i0}+\sigma_{ij})K_{vX_i}\right|\\
  &\quad=\E_{X,\sigma}\sum_vp_v\max_j
    \left|\sum_i(\sigma_{i0}+\sigma_{ij})K_{vX_i}\right|
   \le2H\sqrt{2k\log(2r)}.
\end{aligned}
\end{equation}

\paragraph[Bound the sum weighted by P_X.]{Bound the sum weighted by \(P_X\).}
We can write the \(P_X\)-weighted sum as
\[
  \sum_vP_X(v)\max_j
    \left|\sum_i(\sigma_{i0}+\sigma_{ij})K_{vX_i}\right|
  =\frac1k\sum_{\ell=1}^k\max_j
    \left|\sum_i(\sigma_{i0}+\sigma_{ij})K_{X_\ell X_i}\right|.
\]
Fix \(\ell\) and condition on \(X_\ell=u\). Since \(K_{uu}=0\), we have
\[
  \sum_i(\sigma_{i0}+\sigma_{ij})K_{X_\ell X_i}
  =\sum_{i\ne\ell}(\sigma_{i0}+\sigma_{ij})K_{uX_i}.
\]
The other \(X_i\) are independent draws according to \(p\), so applying
Khintchine, Massart, and Jensen again gives
\[
\begin{aligned}
  &\E_{X,\sigma}\left[
    \max_j\left|\sum_i(\sigma_{i0}+\sigma_{ij})K_{X_\ell X_i}\right|
    \mathrel{\Big|}X_\ell=u\right]\\
  &\quad\le2\sqrt{2\log(2r)}\,
    \E_X\left[
      \left(\sum_{i\ne\ell}K_{uX_i}^2\right)^{1/2}
      \mathrel{\Big|}X_\ell=u\right]\\
  &\quad\le2\sqrt{2\log(2r)}\left((k-1)h_u^2\right)^{1/2}
   =2\sqrt{2(k-1)\log(2r)}\,h_u.
\end{aligned}
\]
We remove the conditioning on \(X_\ell=u\) by the law of total expectation. Summing the
resulting bounds over \(\ell\), dividing by \(k\), and using \(H=\sum_up_uh_u\), we
obtain
\begin{equation}\label{eq:PX-term}
\begin{aligned}
  &\E_{X,\sigma}\sum_vP_X(v)\max_j
    \left|\sum_i(\sigma_{i0}+\sigma_{ij})K_{vX_i}\right|
   \le2\sqrt{2(k-1)\log(2r)}\sum_up_uh_u\\
  &\quad=2H\sqrt{2(k-1)\log(2r)}
   \le2H\sqrt{2k\log(2r)}.
\end{aligned}
\end{equation}

Each of the three unnormalized sums has expected value at most \(2H\sqrt{2k\log(2r)}\).
Substituting these bounds into \eqref{eq:rway-three-weight-split} and dividing by
\(k\) gives \(6\sqrt2\,H\sqrt{\log(2r)/k}\), proving Lemma~\ref{lem:signed-maximum}.

\end{proof}

\subsection*{Putting it together}

\begin{corollary}[Expected error in \(F_{K,r}\)]
\label{cor:functional-error}
Let \(X_1,\ldots,X_k\) be independent draws from \(p\), and let \(P_X\) be their
empirical distribution. Then
\begin{equation}\label{eq:functional-error}
  \E_X|F_{K,r}(P_X)-F_{K,r}(p)|
  \le48\sqrt2\,H\sqrt{\frac{\log(2r)}k}.
\end{equation}
\end{corollary}

\begin{proof}
Taking expectations in Lemma~\ref{lem:rway-value-stability} with \(q=p\) and \(s=P_X\),
and using Lemmas~\ref{lem:linear-reduction} and~\ref{lem:signed-maximum}, gives
\[
  \E_X|F_{K,r}(P_X)-F_{K,r}(p)|
  \le\E_X\Delta_{K,r}(p,P_X)
  \le8\cdot6\sqrt2\,H\sqrt{\frac{\log(2r)}k}
  =48\sqrt2\,H\sqrt{\frac{\log(2r)}k}.
\]
This also proves the sampling bound \eqref{eq:sampling-target}.
\end{proof}

\begin{proof}[Proof of Theorem~\ref{thm:rway-main}]
Lemma~\ref{lem:two-values} identifies the normalized cut-value error with
\(|F_{K,r}(P_X)-F_{K,r}(p)|\). Applying Corollary~\ref{cor:functional-error} proves the
theorem.
\end{proof}

\appendix

\section{Technical lemmas}
\label{app:technical-lemmas}

\subsection{Ledoux--Talagrand contraction}
\label{sec:standard-results}

We use the following standard form of the Ledoux--Talagrand contraction in
Lemma~\ref{lem:adaptive-symmetrization}; see \cite{ledoux-talagrand}.

\begin{lemma}[Absolute-value Rademacher contraction]
\label{lem:absolute-rademacher-contraction}
Let \(\mathcal T\subseteq\mathbb R^k\) be nonempty and compact, and let
\(g:\mathbb R\to\mathbb R\) be \(1\)-Lipschitz with \(g(0)=0\). If
\(\sigma_1,\ldots,\sigma_k\) are independent uniform signs, then
\[
  \E_\sigma\max_{t\in\mathcal T}
  \left|\frac1k\sum_{i=1}^k\sigma_i g(t_i)\right|
  \le
  2\E_\sigma\max_{t\in\mathcal T}
  \left|\frac1k\sum_{i=1}^k\sigma_i t_i\right|.
\]
\end{lemma}

\subsection{Khintchine inequality}
\label{sec:khintchine}

We use the following form of the upper \(L^1\) Khintchine inequality in
Lemma~\ref{lem:pooled-box-width}; see \cite{vershynin}.

\begin{lemma}[Upper \(L^1\) Khintchine inequality]
\label{lem:upper-l1-khintchine}
Let \(k\ge1\) be an integer, let \(a_1,\ldots,a_k\in\mathbb R\), and let
\(\sigma_1,\ldots,\sigma_k\in\{\pm1\}\) be independent uniform signs. Then
\[
  \E_\sigma\left|\sum_{i=1}^k\sigma_i a_i\right|
  \le
  \left(\sum_{i=1}^k a_i^2\right)^{1/2}.
\]
\end{lemma}

\begin{proof}
This \(L^1\) bound follows immediately from \(\E|Z|\le(\E Z^2)^{1/2}\), applied to
\(Z=\sum_i\sigma_i a_i\), since independence and zero means give
\(\E Z^2=\sum_i a_i^2\).
\end{proof}

\subsection{Vector contraction}
\label{app:contraction}

We use the following absolute-value consequence of Maurer's vector contraction
inequality in Lemma~\ref{lem:linear-reduction}; see \cite[Corollary~1]{maurer}.

\begin{lemma}[Absolute-value consequence of Maurer's inequality]
\label{lem:absolute-vector-contraction}
Let \(\mathcal T\subseteq(\R^m)^k\) be compact with \(0\in\mathcal T\), and write
\(t=(t_1,\ldots,t_k)\), where \(t_i=(t_{i1},\ldots,t_{im})\in\R^m\). Let
\(\phi:\R^m\to\R\) be \(L_0\)-Lipschitz in Euclidean norm, with \(\phi(0)=0\). If
\(\tau_i\) and \(\sigma_{i\ell}\), for \(1\le i\le k\) and \(1\le\ell\le m\), are
independent uniform signs, then
\begin{equation}\label{eq:absolute-contraction}
  \E_\tau\max_{t\in\mathcal T}
    \left|\frac1k\sum_{i=1}^k\tau_i\phi(t_i)\right|
  \le2\sqrt2 L_0\,
    \E_\sigma\max_{t\in\mathcal T}
      \left|\frac1k\sum_{i=1}^k\sum_{\ell=1}^m
        \sigma_{i\ell}t_{i\ell}\right|.
\end{equation}
\end{lemma}

\begin{proof}
The finite-dimensional case of \cite[Corollary~1]{maurer}, applied to \(f_t(i)=t_i\) and
\(h_i=\phi\) and divided by \(k\), gives
\[
  \E_\tau\max_{t\in\mathcal T}\frac1k\sum_{i=1}^k\tau_i\phi(t_i)
  \le\sqrt2 L_0\,
    \E_\sigma\max_{t\in\mathcal T}\frac1k
      \sum_{i=1}^k\sum_{\ell=1}^m\sigma_{i\ell}t_{i\ell}.
\]
For each sign vector \(\tau\), the maximum of the signed sum and the maximum of its
negative are both nonnegative. Indeed, \(0\in\mathcal T\) and \(\phi(0)=0\), so choosing
\(t=0\) gives zero for both sums. Their sum is therefore at least the maximum absolute value. Since
\(\tau\) and \(-\tau\) have the same distribution, the two maxima have equal
expectations, so
\[
  \E_\tau\max_{t\in\mathcal T}
    \left|\frac1k\sum_{i=1}^k\tau_i\phi(t_i)\right|
  \le2\E_\tau\max_{t\in\mathcal T}
    \frac1k\sum_{i=1}^k\tau_i\phi(t_i).
\]
Combining these bounds and inserting an absolute value in the linear sum proves the
claim.
\end{proof}

\paragraph[Application to r-way cuts.]{Application to \(r\)-way cuts.}
The function \(\phi(z_0,z_1,\ldots,z_r)=z_0-\min_{1\le j\le r}z_j\) used in
Lemma~\ref{lem:linear-reduction} vanishes at zero and satisfies
\[
\begin{aligned}
  |\phi(z)-\phi(w)|
  &\le |z_0-w_0|+\max_{1\le j\le r}|z_j-w_j|\\
  &\le |z_0-w_0|+
    \left(\sum_{j=1}^r|z_j-w_j|^2\right)^{1/2}
   \le\sqrt2\,\|z-w\|_2.
\end{aligned}
\]
To apply Lemma~\ref{lem:absolute-vector-contraction} to the function \(g_c\), recall that
\(g_c(v)=\phi(f_c(v))\), where
\[
  f_c(v)=\left(\sum_{\ell=1}^r(Kc^\ell)_v,
    (Kc^1)_v,\ldots,(Kc^r)_v\right).
\]
Coordinate \(0\) stores the sum of all part contributions, while coordinate \(j\)
stores the contribution \((Kc^j)_v\) of part \(j\). Keeping the sum as a separate
coordinate gives the Lipschitz constant \(\sqrt2\) above, independent of \(r\).

Apply the lemma with \(m=r+1\), indexing the coordinates and their independent signs
by \(0,\ldots,r\). For each sample index \(i\), its linear signed sum is
\[
\begin{aligned}
  \sum_{j=0}^r\sigma_{ij}\bigl(f_c(X_i)\bigr)_j
  &=\sigma_{i0}\sum_{j=1}^r(Kc^j)_{X_i}
    +\sum_{j=1}^r\sigma_{ij}(Kc^j)_{X_i}\\
  &=\sum_{j=1}^r(\sigma_{i0}+\sigma_{ij})(Kc^j)_{X_i}.
\end{aligned}
\]
Thus each part contribution receives the sign \(\sigma_{i0}\) from the sum coordinate
and the sign \(\sigma_{ij}\) from its own coordinate. Substituting this identity and
\(L_0=\sqrt2\) into Lemma~\ref{lem:absolute-vector-contraction} gives
\eqref{eq:conditional-linear-reduction}, with factor \(2\sqrt2 L_0=4\).

\subsection{Massart's finite-class lemma}
\label{app:signed-maximum}

We use the following form of Massart's finite-class lemma in
Lemma~\ref{lem:signed-maximum}; see \cite[Theorem~3.7]{mohri-rostamizadeh-talwalkar}.

\begin{lemma}[Absolute-value consequence of Massart's lemma]
\label{lem:massart-finite-class}
Let \(m,r\ge1\) be integers, let \(a_{ij}\in\R\), and let \(\sigma_{ij}\in\{\pm1\}\),
for \(1\le i\le m\) and \(1\le j\le r\), be independent uniform signs. Then
\[
  \E_\sigma\max_{1\le j\le r}
    \left|\sum_{i=1}^m\sigma_{ij}a_{ij}\right|
  \le\sqrt{2\log(2r)}\,
    \max_{1\le j\le r}\left(\sum_{i=1}^m a_{ij}^2\right)^{1/2}.
\]
\end{lemma}

\begin{proof}
Set \(\mathcal T:=\{\pm b^j:1\le j\le r\}\), where \(b^j\in\R^{mr}\) has coordinates
indexed by pairs \((i,\ell)\), with \(b^j_{ij}=a_{ij}\) and \(b^j_{i\ell}=0\) for
\(\ell\ne j\). Applying \cite[Theorem~3.7]{mohri-rostamizadeh-talwalkar} to
\(\mathcal T\), after canceling its factor \(1/(mr)\), gives
\[
\begin{aligned}
  &\E_\sigma\max_{1\le j\le r}
    \left|\sum_{i=1}^m\sigma_{ij}a_{ij}\right|\\
  &\quad=\E_\sigma\max_{t\in\mathcal T}
    \sum_{i=1}^m\sum_{\ell=1}^r\sigma_{i\ell}t_{i\ell}
  \le\sqrt{2\log|\mathcal T|}\,\max_{t\in\mathcal T}\|t\|_2\\
  &\quad\le\sqrt{2\log(2r)}\,
    \max_{1\le j\le r}\left(\sum_{i=1}^m a_{ij}^2\right)^{1/2}.
\end{aligned}
\]
Here \(|\mathcal T|\le2r\) and \(\|b^j\|_2^2=\sum_i a_{ij}^2\).
\end{proof}

\section*{AI disclosure}
We used ChatGPT as a writing assistant to improve the writing and clarity of the
paper, and Perplexity to help identify relevant literature we might have missed.
The results and proof ideas were developed by the authors, who checked all
AI-assisted content and are responsible for the paper's accuracy and originality.

\bibliographystyle{alphaurl}
\bibliography{references}

\end{document}